%% file: 0.head.tex
\documentclass[conference]{IEEEtran}
\usepackage{cite}
\usepackage{amsmath,amssymb,amsfonts}
\usepackage{algorithmic}
\usepackage{graphicx}
\usepackage{xcolor}
\usepackage{makecell}
\usepackage{booktabs} 
\usepackage{subfig}
\usepackage{amsmath}
\usepackage[normalem]{ulem}
\usepackage{enumitem}
\usepackage{multirow}
\usepackage[nomargin,inline,marginclue,draft]{fixme}
\usepackage{balance}
\usepackage{changepage}
\usepackage[colorlinks,linkcolor=blue]{hyperref}

\newcommand{\rev}[1]{\textcolor{black}{#1}}

\newlength\savedwidth

\def\BibTeX{{\rm B\kern-.05em{\sc i\kern-.025em b}\kern-.08em
		T\kern-.1667em\lower.7ex\hbox{E}\kern-.125emX}}
\begin{document}
	
	\title{Group-Buying Recommendation for Social E-Commerce}

	\author{\IEEEauthorblockN{Jun~Zhang,
			Chen~Gao\textsuperscript{\textsection}, 
			Depeng~Jin,
			Yong~Li
		}
		\IEEEauthorblockA{Beijing National Research Center for Information Science and Technology\\
			Department of Electronic Engineering, Tsinghua University, Beijing 100084, China\\}
		zhangjun990222@gmail.com, gc16@mails.tsinghua.edu.cn, \{jindp, liyong07\}@tsinghua.edu.cn
	}
	\maketitle

	\begingroup\renewcommand\thefootnote{\textsection}
	\footnotetext{Chen Gao is the Corresponding Author.}
	\endgroup
	
	\begin{abstract}
		Group buying, as an emerging form of purchase in social e-commerce websites, such as Pinduoduo\footnote{https://www.pinduoduo.com/}, has recently achieved great success. In this new business model, users, \textit{initiator}, can launch a group and share products to their social networks, and when there are enough friends, \textit{participants}, join it, the deal is clinched. Group-buying recommendation for social e-commerce, which recommends an item list when users want to launch a group, plays an important role in the group success ratio and sales. However, designing a personalized recommendation model for group buying is an entirely new problem that is seldom explored. 
		In this work, we take the first step to approach the problem of group-buying recommendation for social e-commerce and develop a GBGCN method (short for Group-Buying Graph Convolutional Network). Considering there are multiple types of behaviors (launch and join) and structured social network data, we first propose to construct directed heterogeneous graphs to represent behavioral data and social networks. We then develop a graph convolutional network model with multi-view embedding propagation, which can extract the complicated high-order graph structure to learn the embeddings. Last, since a failed group-buying implies rich preferences of the initiator and participants, we design a double-pairwise loss function to distill such preference signals.
		We collect a real-world dataset of group-buying and conduct experiments to evaluate the performance. Empirical results demonstrate that our proposed GBGCN can significantly outperform baseline methods by 2.69\%-\rev{7.36\%}. \rev{The codes and the dataset are released at \url{https://github.com/Sweetnow/group-buying-recommendation}.}
	\end{abstract}
	
	
	\input{1.intro.tex}

	\input{2.probdef.tex}

\input{3.method.tex}
	\input{4.exp.tex}

\input{5.related.tex}
	\input{6.conclusion.tex}

\section*{Acknowledgement}
This work was supported in part by The National Key Research and Development Program of China under grant 2018YFB1800804, the National Nature Science Foundation of China under U1936217, 61971267, 61972223, 61941117, 61861136003, Beijing Natural Science Foundation under L182038, Beijing National Research Center for Information Science and Technology under 20031887521, and research fund of Tsinghua University - Tencent Joint Laboratory for Internet Innovation Technology.
	
	\balance
	\bibliographystyle{IEEEtran}
	\bibliography{bibliography}

\end{document}

%% file: 1.intro.tex
\section{Introduction}\label{sec::intro}

\begin{figure}[t]
	\begin{center}
		\mbox{
			\subfloat{\includegraphics[width=0.7\linewidth]{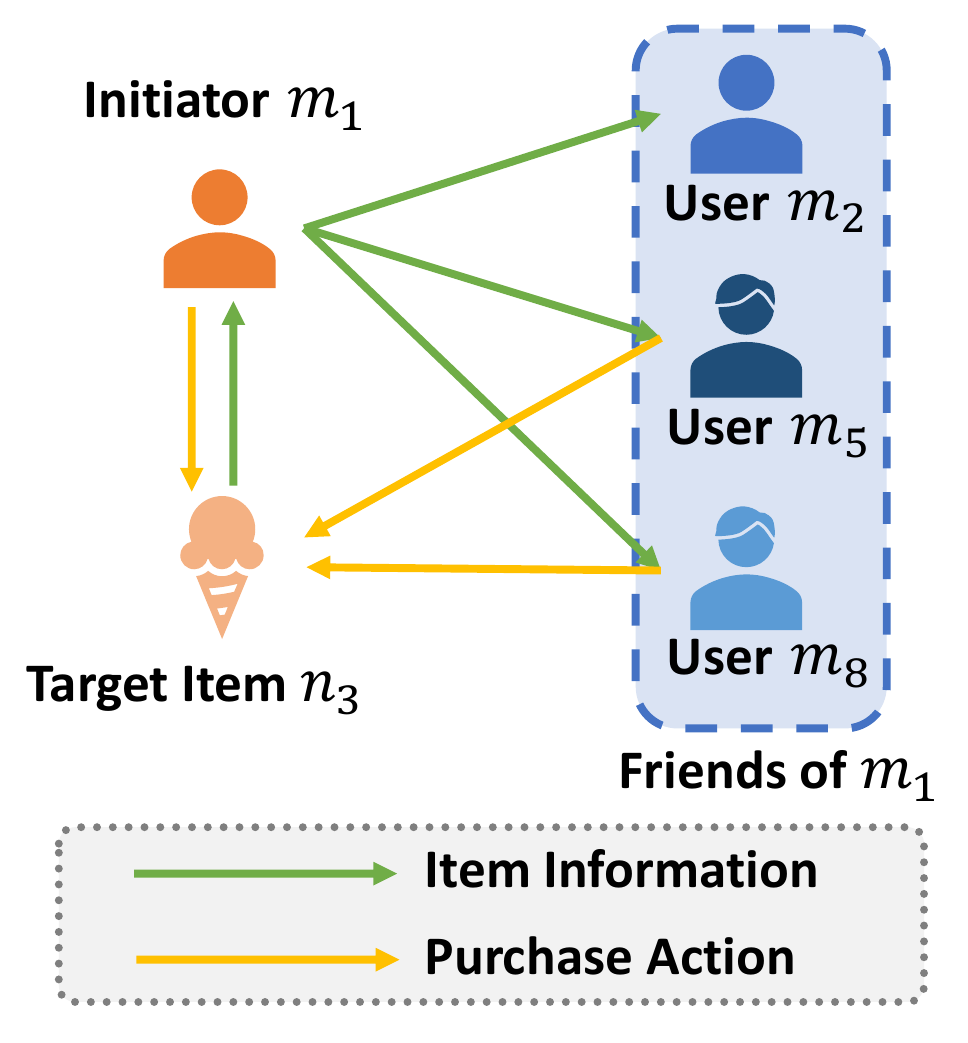}}
		}
	\end{center}
	\caption{The illustration of the process of one user launching a group buying, sharing the target item to friends and clinching the group buying.}
	\label{fig::gb}
\end{figure}

Group buying, as a new form of business model in social e-commerce websites, such as Pinduoduo, becomes increasingly popular. It was reported by Forbes\footnote{https://www.forbes.com/sites/eladnatanson/2019/12/04/the-miraculous-rise-of-pinduoduo-and-its-lessons} that group-buying has achieved great access in attracting new users, increasing user activities, improving platform profits, etc. Specifically, in this purchase model, a user can share the product, which he/she is interested in, broadcasting to his/her friends in social media. This stage is called \textit{lanuching} a group and the user is called \textit{initiator}. If one or several friend(s) who also want(s) to purchase the shared product, he/she (they) can take participate in the group. This stage is called \textit{joining} a group and such friend(s) is (are) called \textit{participant(s)}. When there are enough users in this group, group buying is clinched. This stage is called \textit{clinching} a group. The whole process described above is illustrated in Figure~\ref{fig::gb}.
In short, a group buying can be understood that a group of users interact with a product, through the initiator's sharing and participants' following.

The problem of \textbf{group-buying recommendation \rev{for social e-commerce}} is defined to recommend product-candidates for the initiator to launch a group buying. With suitable recommended products customized for the initiator, the preferences of the initiator and the participants are more likely to be satisfied, and the group buying becomes more probable to clinch. Thus, it is essential and beneficial for the social e-commerce website to design powerful models for the group-buying recommendation. However, this is a completely new problem that is rarely explored by existing works. Whether a group succeeds to clinch or not is determined by multiple factors, including the initiator's interest, participants' interest, initiator's influence on other users, etc.

In fact, there are two kinds of recommendation tasks related to the problem of group-buying recommendation in social e-commence: social recommendation~\cite{jamali2010matrix,guo2015social,soreg,soregbpr,wang2017item,ste,mtrust,icde-ensemble,sorec,localbal,DiffNet,fan2019graph} and group recommendation~\cite{baltrunas2010group,liu2012exploring,quintarelli2016recommending,yuan2014generative,christensen2016social,hu2014deep,cao2018attentive,cao2019social,yin2019social}. Social recommendation is defined as utilizing social-relational data to enhance recommendation. A commonly accepted paradigm is to make use of friends' preference to help model users, assuming friends tend to have similar tastes. However, in the problem of group-buying recommendation, the form of users interacting with products is completely different, where the launching and joining behaviors cannot be handled. On the other hand, group recommendation is defined as to recommend one product to a group. For example, in location-based services, the platform can recommend a travel destination to a family. However, in group recommendation, the group is always fixed (a family, for example). While in the group buying, the group is dynamic and a sub-sample from the entire social network, and how to determine the potential group of users is important. Therefore, existing solutions of both social recommendation and group recommendation cannot serve the task of group-buying recommendation.

Generally, the problem of group-buying recommendation for social e-commerce is faced with the following three major challenges:
\begin{itemize}[leftmargin=*]
	\item \textbf{Multiple roles of users in group buying.} In our problem, users can both launch a group or join a friend's group. In other words, users can play as different roles, initiators or participants, with different interests and motivations, making it challenging to capture accurately.
	
	\item \textbf{Complicated social influence.} Despite user interests, the initiator's influence on the social network is another significant factor determining whether the friend joins. It is challenging to model social influence and user interests jointly.

	\item \textbf{\rev{Complicated feedback of group-buying interaction data.}} 
	\rev{In our problem, the predictive signal is revealed by interaction data, since it is related to many users' preferences. Thus, it is challenging to learn from such complicated feedback. }
\end{itemize}

To overcome the above mentioned challenges in our problem, we propose a 
solution named GBGCN (short for \textbf{G}roup-\textbf{B}uying \textbf{G}raph \textbf{C}onvolutional \textbf{N}etworks).
Our GBGCN approaches the problem via graph convolutional network-based representation learning on the graph-structured group-buying data.
Precisely, to capture the multiple roles of users in our problem, we develop in-view user-item embedding propagation to extract the preferences of different roles.
We then employ cross-view embedding propagation for learning social influence and generate comprehensive representation. Last, we design a double-pairwise loss function to distill predictive signals that are related to many users in a group.

To summarize, the main contributions of this work are as follows,

\begin{itemize}[leftmargin=*]
	\item To the best of our knowledge, we are the first to approach the problem of personalized recommendation for group buying \rev{in social e-commerce}, which is significant but not explored yet.
	\item We develop a graph convolutional network (GCN) based method with embedding propagations to model multiple user roles and social influence and a double-pairwise loss function to utilize the complex feedback in our problem.
	\item We collect a precious dataset from a large-scale social e-commerce website and conduct extensive experiments to evaluate the performance of GBGCN. Empirical results demonstrate that our GBGCN can help improve the recommendation performance significantly. 
\end{itemize}

The remainder of the paper is as follows. We first formalize
the problem in Section~\ref{sec::probdef} and present our proposed method in Section~\ref{sec::method}. We then
conduct experiments in Section~\ref{sec::exp} and review related
work in Section~\ref{sec::related}. Last, we conclude the paper in Section~\ref{sec:conclusion}.

%% file: 2.probdef.tex
\section{Problem Formulation}\label{sec::probdef}

\begin{table}[t]
	\centering
	\caption{A list of commonly used notations.}
	\begin{tabular}{c|l}
		\hline
		\textbf{Notat.} & \textbf{Description}\\ \hline
		$\mathcal{M}$ & The set of users. \\ \hline
		$\mathcal{N}$ & The set of items. \\ \hline
		$P$ & The number of users.\\ \hline
		$Q$ & The number of items. \\ \hline
		$m$ & \makecell[l]{User's ID. (The subscript $i$($p$) if existed \\ means the user is in initiator view (participant view).)}\\ \hline
		$n$ & Item's ID. (The subscript is the same as $m$.) \\ \hline
		$M_p$ & Participant set. \\ \hline
		$t_n$ & The threshold of group size for item $n$.\\ \hline
		$b$ & One group-buying behavior. \\ \hline
		$B$ & The group-buying behavior set.\\ \hline
		$\mathbf{S}$ & The social network.\\ \hline
		$\sigma(\cdot)$ & The activation function. \\ \hline
		$\mu(\cdot) $ & Sigmoid function. \\ \hline
		$\alpha$ & The role coefficient.\\ \hline
		$\beta$ & The loss coefficient.\\ \hline
		$W$ & \makecell[l]{The transformation matrix. (The first subscript \\ denotes source and the second denotes target.)} \\ \hline
		$b$ & \makecell[l]{The transformation bias. \\(The subscripts are the same as $W$.)} \\ \hline
		$\mathcal{L}$ & The fine-grained loss function. \\ \hline
		$\mathcal{L_{-(+)}}$ & The loss function for failed (successful) behaviors.\\ \hline
		$\mathcal{G}$ & The set of the directed heterogeneous graphs.\\ \hline
		$G_{i(p,s)}$ & \makecell[l]{The directed initiator-item \\(participant-item, initiator-participant) interaction graph.}\\ \hline
		$m_{p_j}$ & The j-th participant. \\ \hline
		$\mathcal{N}_s^{I(O)}(\cdot)$ & \makecell[l]{A function that mapping an input vertex \\to its incoming (outgoing) neighborhood in $G_s$.} \\ \hline
		$\mathcal{N}_{i(p)}(\cdot)$ & \makecell[l]{A function that mapping an input vertex \\to its neighborhood in $G_{i(p)}$.}\\ \hline
		$d$ & The embedding size. \\ \hline
		$u_m$ & \makecell[l]{The embedding of user m. \\(The second subscript if existed means which view \\ it belongs to and the superscipt \\ if existed means which layer generates it.)}\\ \hline
		$v_n$ & \makecell[l]{The embedding of item n.\\(The subscript and superscipt are the same as $u_m$.)}\\ \hline
		$L$ & The number of layers.\\ \hline
		 $l$ & The l-th layer.\\ \hline
		$\cdot||\cdot$ & Concatenation of two vectors. \\ \hline
		$y_{mn}$ & The prediction score of user m with item n.\\ \hline
		$B_{-(+)}$ & The failed (successful) part of $B$. \\ \hline

\hline
	\end{tabular}
	\label{tab::symbal}
\end{table}

In social e-commerce, users can share items in her/his social networks, and group buying becomes a remarkable business model. In this model, users can launch a group buying and share it to online social networks. When there are friends taking participants in the group, the group-buying is clinched, or in other words, \textit{succeeded}. If there are no or not enough friends to join, the group buying will be called a \textit{failed} one.

Thus, users have two kinds of roles in group buying, \textit{initiators}, who launch the group buying and sharing products to social networks and \textit{participants} who join the group buying launched by his/her friends. The product, which is the target of the group buying, can be called \textit{target item}.
The e-commerce platform needs to provide recommendations to help users choose the proper target item to launch the group buying as an initiator. Let $\mathcal{M}$ and $\mathcal{N}$ denote the sets of users and items of which the sizes are $P$ and $Q$, respectively.

The behavioral records of group buying are denoted as $B$ consisting of triads of which one record is denoted as $b = \langle m_i, n, M_p \rangle$, where $m_i$ denote initiator user, $n$ denotes the target item, and the $M_p$ denotes the set of participants. \rev{The platform sets a threshold $t_n$ for each item when the item is launched, which is always much less than the number of user's friends,} and the group buying can be called a failed one if $|M_p|$ is less than $t_n$. Note that users in $M_p$ are from the online social network of the initiator user $m_i$. Here we use $\mathbf{S}$, a $P \times P$ symmetric binary matrix to denote the social network. If user $m$ and $m'$ are friends in online social network, then $S_{mm'}=1$ else $S_{mm'}=0$.

With the above background and notations, we can formulate the problem of group-buying recommendation for social e-commerce as follows:

\textbf{Input}: Group-buying behaviors $B$ and online social relations $\mathbf{S}$.

\textbf{Output}: \rev{A model that calculates a score for representing the unnormalized probability of user $m$ launching a \textit{successful} group to buy item $n$.}

After obtaining the model, for a given user $m_i$, we can utilize the model to score all items and select the top-ranked items as the recommendation results for $m_i$.

%% file: 3.method.tex
\section{Methodology}\label{sec::method}

\begin{figure*}[t]
	\begin{center}
		\mbox{
			\subfloat{\includegraphics[width=0.85\linewidth]{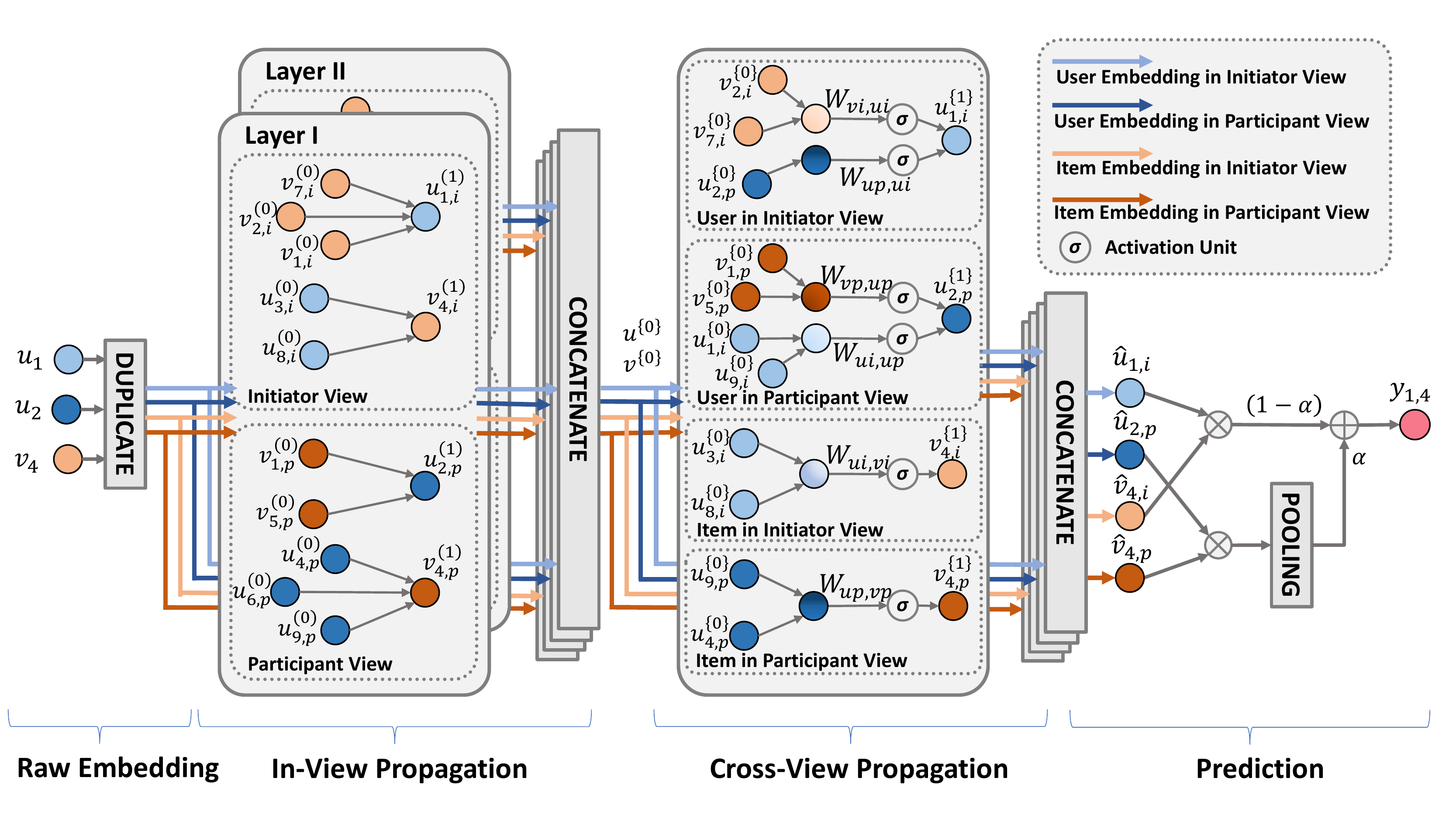}}
		}
	\end{center}
	\vspace{-0.3cm}
	\caption{The architecture of our proposed GBGCN model which is formed by cascading raw embedding layer, in-view propagation, cross-view propagation, and prediction. Here we take the prediction of item 4 and user 1 with only one friend user 2 as an example; as an initiator, user 1 has interacted with item 1, 2, 7; as a participant, user 2 has interacted with item 1, 5; item 4 has been interacted with user 3, 8 as initiators and with user 4, 6, 9 as participants. Some details, such as bias in the cross-view propagation part, are ignored due to space limitations. (Best view in color)} 
	\label{fig::framework}
\end{figure*}

To solve the problem of group-buying recommendation for social e-commerce and overcome the three major challenges mentioned above, we organize multiple roles and complex interactions into directed heterogeneous graphs and introduce multi-graph convolution networks to obtain precise embedding representation from the graphs. We name our proposed solution GBGCN. Figure~\ref{fig::framework} illustrates the architecture of our GBGCN model, which is made up of the four parts as follows.
\begin{itemize}[leftmargin=*]
	\item \textbf{Raw Embedding Layer.}
	This layer initializes embeddings for user and item as raw representation regardless of roles.
	\item \textbf{In-View Propagation.} 
	To distinguish user interests and item properties, we divide the group-buying behaviors into initiator-item interactions and participant-item interactions.
	After this, we refine better representation, considering the two roles by propagating information on the two user-item interaction bipartite graphs.
	\item \textbf{Cross-View Propagation.}
	In this layer, we use social interactions, sharing behaviors and joining behaviors, to guide the information propagation between different roles and model user influence. 
	User-item interactions are utilized again for helping generate exhaustive representation. 
	Fully connected layers (FC) are applied to transform information from source subspaces to target subspaces and extract the meaning parts.
	\item \textbf{Prediction.}
	Finally, we adopt the inner product to capture user's interests for the given item according to his/her role.
	And then, the layer outputs the combination of the initiator's preference and his/her friends' preference as the score of launching a successful group buying.
\end{itemize}

\subsection{Constructing Directed Heterogeneous Graphs}

\begin{figure}[t]
	\begin{center}
		\mbox{
			\subfloat{\includegraphics[width=0.90\linewidth]{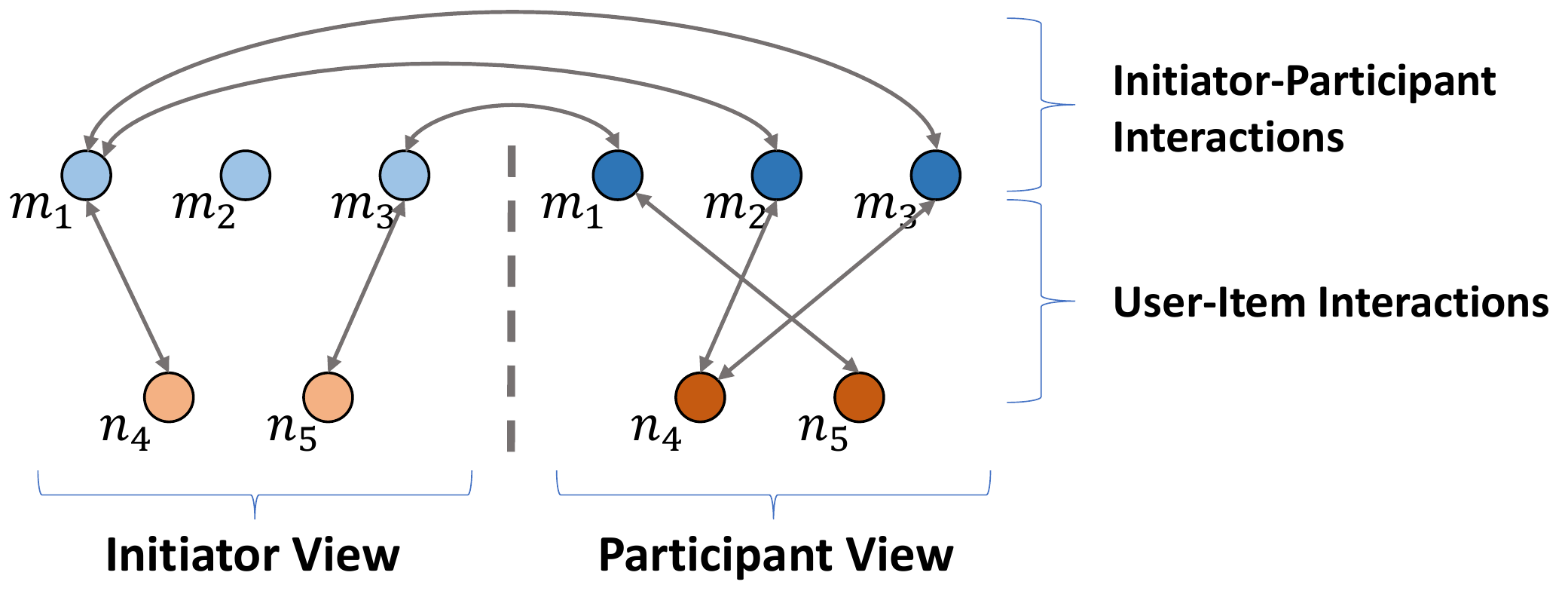}}
		}
	\end{center}
	\caption{The illustration of the heterogeneous graphs with two views and multiplex interactions to describe group buying.} 
	\label{fig::view}
\end{figure}

In order to utilize all categories of behaviors, directed heterogeneous graphs are built to represent the two roles and multiplex interactions in our problem.
We firstly introduce the concept of \textit{views}.
In our problem, both launching and joining can be regarded as user-item interactions. However, users may have potentially different interests when launching and joining.
That is, an initiator should consider his/her friends' interests before launching and a participant should consider not only his/her own interests but also whether the relationship is close or not before joining.
Therefore, we treat the whole user-item interactions as two parts, initiator-item interactions and participant-item interactions. 
Meanwhile, the graphs for organizing all behaviors consist of two views, \textit{initiator view} that contains initiator-item interactions and \textit{participant view} that contains participant-item interactions.
Besides, we consider initiator-participant interactions, sharing, as cross-view connections. These settings are illustrated in Figure~\ref{fig::view}.

Following the settings above, we define the heterogeneous graphs as $\mathcal{G}=\{G_{i}, G_{p}, G_{s}\}$, where $G_{i}$, $G_{p}$, $G_{s}$ denote the initiator view, the participant view and the sharing relations, respectively.
Vertices in $G_{i}$ and $G_{p}$ both consist of user nodes $m\in \mathcal{M}$ and item nodes $n \in \mathcal{N}$. Vertices in $G_{s}$ consist of user nodes $m\in \mathcal{M}$.
For directed edges in graphs, considering a group-buying behavior $b=\langle m_i, n, M_p \rangle$, where $M_p=\{m_{p_1}, m_{p_2}, \cdots, m_{p_{|M_p|}}\}$,  $G_{i}$ contains a bidirectional edge $(m_i, n)$, $G_{p}$ contains $|M_p|$ bidirectional edges $(m_{p_j},n), j=1,2,\cdots,|M_p|$ and $G_{s}$ contains $|M_p|$ directed edges  $(m_i, m_{p_j}), j=1,2,\cdots,|M_p|$ from initiator to participants.
It is noted that we will use incoming neighborhood $\mathcal{N}^{I}_s$ and outgoing neighborhood $\mathcal{N}^{O}_s$ of $G_s$ to distinguish shared and sharing behaviors in cross-view propagation below.

\subsection{Multi Graph Convolution Networks}
\subsubsection{Raw Embedding Layer}

Similar to the existing graph-based recommendation works~\cite{wang2019neural,wang2019kgat,ying2018graph}, we represent each user/item ID with a vector representation as known as embedding to characterize the latent features.
Formally, we represent user $m$ and item $n$ with embedding $u_m\in\mathbb{R}^d$ and $v_n\in\mathbb{R}^d$, where $d$ is the embedding size.

\subsubsection{In-View Propagation}

It is found by existing researches on graph-based recommeders~\cite{NGCF,he2020lightgcn} that message propagation on the user-item bipartite graph is able to explicitly utilize collaborative signals and distill useful information efficiently from user-item interactions.
Intuitively, user's first-order neighbors on the user-item interaction graph, the purchased items represent the user's preference directly, and the second-order neighbors are often considered to have similar interests.
In our problem, user-item interaction graphs are divided into two parts, initiator view and participant view. Each view carries the complete interaction records under the corresponding user role.
Inspired by these, we devise graph convolution layers without FC layers following~\cite{he2020lightgcn} to perform the embedding process on the two views and generate the role-specific embeddings from the raw embeddings.

The process in initiator view can be fomulated as follows:
\begin{equation}\label{eqn::inview-init}
\begin{split}
u_{m,i}^{(l)}&=\frac{1}{|\mathcal{N}_i(m)|}\sum_{n'\in\mathcal{N}_i(m)}v_{n',i}^{(l-1)}, \\
v_{n,i}^{(l)}&=\frac{1}{|\mathcal{N}_i(n)|}\sum_{m'\in\mathcal{N}_i(n)}u_{m',i}^{(l-1)}, \\
where &\quad u_{m,i}^{(0)}=u_m,\quad v_{n,i}^{(0)}=v_n,
\end{split}
\end{equation}
where $l$ denotes the number of layers and $\mathcal{N}_i(\cdot)$ maps one vertex to its neighborhood in $G_i$. It should be noted that the second subscript $i$ in $u_{m,i}$ and $v_{n,i}$ means that the embeddings are under initiator view.

It is almost consistent that the process in participant is fomulated as follows:
\begin{equation}\label{eqn::inview-prtc}
\begin{split}
u_{m,p}^{(l)}&=\frac{1}{|\mathcal{N}_p(m)|}\sum_{n'\in\mathcal{N}_p(m)}v_{n',p}^{(l-1)}, \\
v_{n,p}^{(l)}&=\frac{1}{|\mathcal{N}_p(n)|}\sum_{m'\in\mathcal{N}_p(n)}u_{m',p}^{(l-1)}, \\
where & \quad u_{m,p}^{(0)}=u_m,\quad v_{n,p}^{(0)}=v_n,
\end{split}
\end{equation}
where $\mathcal{N}_p(\cdot)$ maps one vertex to its neighborhood in $G_p$ and the second subscript $p$ means that those embeddings are under participant view.

After propagating with $L$ layers, we generate multiple embeddings with information from different order neighborhood. To leverage them  effectively and avoid information loss, it is intuitive that concatenate all levels' embeddings is a suitable manner:
\begin{equation}\label{eqn::inview-concat}
\begin{split}
u_{m, i}^{\{0\}} &= u_m || u_{m,i}^{(1)} || \cdots || u_{m,i}^{(L)},\\
v_{n, i}^{\{0\}}& = v_n || v_{n,i}^{(1)} || \cdots || v_{n,i}^{(L)},\\
u_{m, p}^{\{0\}} &= u_m || u_{m,p}^{(1)} || \cdots || u_{m,p}^{(L)},\\
v_{n, p}^{\{0\}} &= v_n || v_{n,p}^{(1)} || \cdots || v_{n,p}^{(L)}.
\end{split}
\end{equation}

In the in-view propagation layers, only one set of raw embeddings participate in the propagation process of both initiator view and participant view rather than two sets.
We argue that, on the one hand, the design avoids inequality of model capacity; on the other hand, it requires the raw embedding to learn the essential features from two kinds of user-item interactions.

\subsubsection{Cross-View Propagation}
After in-view propagation layers, we obtain two sets of embeddings extracting user-item interaction information for initiator and participant view separately.
Our key problem further turns to model the complicated social influence, which is also significant in determining whether a launched group clinches or not.

To solve this problem, we adopt cross-view information propagation with FC layers for extracting social influence and transform embeddings to corresponding subspaces.
In particular, every user node in initiator view propagates self-information to users who used to join the user's group to simulate sharing behaviors and aggerates information in participant view from those users as the simulation of response in participants' joining behaviors at the same time.
To introduce more preference information as a supplement, the characteristics of interacted items are also considered in this layer.
However, it is unreasonable to put the social influence related embeddings and preference related embeddings together.
To overcome it, we adopt FC layers to distill meaningful parts and transform the embeddings from their original subspaces to the target subspace corresponding to the target view.
For item nodes, we adopt the network structure consistent with the above to ensure the correspondence in the subspaces.
We summarize the above design into the following formulas:
\begin{equation}\label{eqn::crossview1}
\begin{split}
u_{m,i}^{\{1\}}&=\sigma\left(\left(\frac{1}{|\mathcal{N}_i(m)|}\sum_{n'\in\mathcal{N}_i(m)}v_{n',i}^{\{0\}}\right)W_{vi,ui}+b_{vi,ui}\right) \\
&+\sigma\left(\left(\frac{1}{|\mathcal{N}^{O}_s(m)|}\sum_{m'\in\mathcal{N}^{O}_s(m)}u_{m',p}^{\{0\}}\right)W_{up,ui}+b_{up,ui}\right),
\end{split}
\end{equation}

\begin{equation}\label{eqn::crossview2}
v_{n,i}^{\{1\}}=\sigma\left(\left(\frac{1}{|\mathcal{N}_i(n)|}\sum_{m'\in\mathcal{N}_i(n)}u_{m',i}^{\{0\}}\right)W_{ui,vi}+b_{ui,vi}\right),
\end{equation}

\begin{equation}\label{eqn::crossview3}
\begin{split}
u_{m,p}^{\{1\}}&=\sigma\left(\left(\frac{1}{|\mathcal{N}_p(m)|}\sum_{n'\in\mathcal{N}_p(m)}v_{n',p}^{\{0\}}\right)W_{vp,up}+b_{vp,up}\right) \\
&+\sigma\left(\left(\frac{1}{|\mathcal{N}^{I}_s(m)|}\sum_{m'\in\mathcal{N}^{I}_s(m)}u_{m',i}^{\{0\}}\right)W_{ui,up}+b_{ui,up}\right),
\end{split}
\end{equation}

\begin{equation}\label{eqn::crossview4}
v_{n,p}^{\{1\}}=\sigma\left(\left(\frac{1}{|\mathcal{N}_p(n)|}\sum_{m'\in\mathcal{N}_p(n)}u_{m',p}^{\{0\}}\right)W_{up,vp}+b_{up,vp}\right),
\end{equation}
where $\sigma(\cdot)$ denotes the activation function, $\mathcal{N}_s^O(\cdot)$ and $\mathcal{N}_s^I(\cdot)$ map one vertex to its outgoing and incoming neighborhood in $G_s$, respectively. 
All $W\in\mathbb{R}^{(L+1)d\times(L+1)d}$ and $b\in\mathbb{R}^{(L+1)d}$ in Equation~\ref{eqn::crossview1}-\ref{eqn::crossview4} are the trainable transformation matrices and biases, respectively.
The subscripts in $W$ and $b$ mean the embeddings' source subspace and the target subspace.
For example,  the subscripts in $W_{up,ui}$ mean that the matrix transforms embeddings from \textbf{u}ser subspace in \textbf{p}articipant view to \textbf{u}ser subspace in \textbf{i}nitiator view.

Following the manner used in Equation~\ref{eqn::inview-concat}, we concatenate the input embeddings with generated embeddings in this layer to obtain final vector representation as follows:
\begin{equation}\label{eqn::crossview-concat}
\begin{split}
\hat{u}_{m, i} = u_{m,i}^{\{0\}} || u_{m,i}^{\{1\}}, \quad\hat{v}_{n, i} = v_{n,i}^{\{0\}} || v_{n,i}^{\{1\}},\\
\hat{u}_{m, p} = u_{m,p}^{\{0\}} || u_{m,p}^{\{1\}}, \quad\hat{v}_{n, p} = v_{n,p}^{\{0\}} || v_{n,p}^{\{1\}}.
\end{split}
\end{equation}

\subsubsection{Prediction}
After obtaining the final embeddings of both user and item in the two views, it is the next problem that how to predict the score of launching a successful group buying and return a ranking list given a user and item candidate set.
We adopt a predictive function to estimate $y_{mn}$ which is the score of user $m$ launching a successful group buying with item $n$ as follows:
\begin{equation}\label{eqn::pred}
y_{mn}=(1-\alpha)\hat{u}_{m,i}^T \hat{v}_{n,i}+\alpha\left(\frac{\sum_{m'=0}^{P-1}S_{mm'}\hat{u}_{m',p}^T \hat{v}_{n,p}}{\sum_{m'=0}^{P-1}S_{mm'}}\right),
\end{equation}
where $\alpha$ is a hyper-parameter that should be tuned.
In this predictive function, we calculate the similarity between the potential initiator $m$ and the target item $n$ in initiator view, and then calculate the average of the similarity between the initiator's friends and the target item in participant view, followed by weighted sum.
The first step is to estimate the unnormalized probability of launching a group buying.
The second step is to estimate whether there are enough participants for a successful deal according to the initiator's friends' interests.
The role coefficient $\alpha$ indicates the importance of initiator and participant in recommeder system. For the weighted sum, we argue that the strong wish of either initiator or participant may lead to a successful deal. We leave the exploration of more complex predictive function for future works.

In conclusion, we address the first two major challenges of our problem by constructing directed heterogeneous graphs, multi-graph convolution networks and the prediction function. The solution to the last challenge will be introduced below.

\subsection{Model Training}
\subsubsection{Fine-grained Loss Function}

The last major challenge of our problem is the complicated feedback revealing fused preference signals of many users.
More specifically, we can observe both successful and failed group-buying behaviors.
A failed behavior shows that the initiator paid for the target item and launched a group buying, but there are not enough friends to join.
Such behaviors imply that the initiators like the target items but their friends are not attracted to the target items to a certain degree, or the social influence between initiators and their friends is relatively weak. For those successful behaviors, it is clear that the initiators and the participants are all interested in the target items.

Based on these prior knowledge, we propose a fine-grained loss function based on \textit{Bayesian Personalized Ranking} (BPR)~\cite{rendle2009bpr} to mine the information in the behaviors better.
For one failed behavior $b=\langle m_i, n, M_p \rangle$ where $|M_p|$ is less than threshold\footnote{\rev{Threshold is a kind of value that highly depends on the dataset. In other words, it is determined by the service provider and cannot be directly modeled. Here our double-pairwise loss can leverage it indirectly.}}
, the loss function is as follows:

\begin{equation}\label{eqn::loss-failed}
\mathcal{L}_{-}(b)=-\text{ln}\mu\left(y_{m_i n}-y_{m_i n'}\right)-\beta\sum_{m'=0}^{P-1}S_{mm'}\text{ln}\mu\left(y_{m' n'}-y_{m' n}\right),
\end{equation}
where $\mu(\cdot)$ denotes sigmoid function, the loss coefficient $\beta$ is a hyper-paramter to control the strength of belief that user $m_i$'s friends dislike the target item $n$ and item $n'$ is an unobserved item (i.e. negative sample by randomly selection).
For one successful behavior $b=\langle m_i, n, M_p \rangle$, the loss function is as follows:
\begin{equation}\label{eqn::loss-successful}
\mathcal{L}_{+}(b)=-\text{ln}\mu\left(y_{m_i n}-y_{m_i n'}\right)-\sum_{m'\in M_p}\text{ln}\mu\left(y_{m' n}-y_{m' n'}\right).
\end{equation}

Considering both failed behaviors $B_{-}$ and successful behaviors $B_{+}$ split from the entire behaviors $B$, the fine-grained loss function is:
\begin{equation}\label{eqn::loss-all}
\mathcal{L}=\sum_{b\in B_{+}}\mathcal{L}_{+}(b) + \sum_{b\in B_{-}}\mathcal{L}_{-}(b).
\end{equation}
Our training goal is to minimize the fine-grained loss function by optimize the model parameters.  Besides, we conduct $L_2$ normalization to avoid over-fitting and social regularization term proposed in~\cite{jamali2010matrix} for better learning.

\subsubsection{Mini-batch Training}

We perform mini-batch training for faster training speed.
For constructing a mini-batch, we first sample a batch of group-buying behaviors $b=\langle m_i, n, M_p \rangle$, and then we adopt negative sampling technique, which is widely used in recommender systems~\cite{ncf,wang2019neural,rendle2009bpr} to handle implicit, \textit{i.e.}, binary, feedback, to randomly select unobserved items $\{n'_1, n'_2, \cdots, n'_k\}$ for each observed behaviors with a sampling ratio of $k$.
After sampling, we obtain $k$ quadruples $\{\langle m_i, n, M_p, n'_1 \rangle, \cdots, \langle m_i, n, M_p,n'_k \rangle\}$ for each instance in the mini-batch. With the constructed mini-batch, an optimization step is taken to minimize our loss function. 

\subsubsection{Pre-training}

Due to the data sparsity problem, it is difficult to train both embeddings and FC layers jointly. ~\cite{ncf} also points out that neural networks are sensitive to initialization. To better train our GBGCN, we follow~\cite{ncf, cao2018attentive} and pre-train its raw embeddings with an extremely simplified version of GBGCN that removes all propagation layers by Adam~\cite{kingma2014adam}. With the normalized pre-trained embeddings as an initialization, we adopt stochastic gradient descent (SGD) to fine-tune the pre-trained model.

\subsection{Discussion}
\rev{The three key components of our GCN-based method can be divided into three parts, which correspond to the three main challenges in the group-buying recommendation for social e-commerce. For the first challenge of user’s multiple-role property, we design the in-view propagation layers for capturing the role-aware user interests; for the second challenge of complicated social influence, we design the cross-view propagation to modeling the social relations; for the third challenges of complex feedback, we design the double-pair loss function.}

%% file: 4.exp.tex
\section{Experiments}\label{sec::exp}
In this section, we first collect a precious real-world group-buying dataset and conduct extensive experiments to answer the following research questions.

\begin{itemize}[leftmargin=*]
	\item RQ1: Can our proposed GBGCN achieve the best recommendation performance compared with baselines?
	\item \rev{RQ2: How time efficient are our GBGCN and baselines?}
	\item RQ3: How do the multi-view design in our GBGCN affect the performance?
	\item RQ4: How do the hyper-parameters affect the performance?
	\item RQ5: What insight can the learned embeddings provide?
\end{itemize}

In the next parts, we describe the experimental settings first, followed by the answers to the four research questions above.

\subsection{Experimental Settings}
\subsubsection{Data Collection and Preprocessing}
Since there is no public group-buying dataset available, we collect the dataset from an e-commerce website Beibei\footnote{https://www.beibei.com}, the largest e-commerce platform for maternal and infant products in China.
\rev{To be more specifically, we obtained group-buying logs from Beibei's production database, simply filtered out users and items with few interactions as a widely-used manner mentioned in ~\cite{he2016fast,he2018outer,rendle2009bpr}, and performed ID remapping to protect user privary.}
\rev{We summarize the statistics of the dataset in Table~\ref{tab::stat} and publish the dataset at \url{https://github.com/Sweetnow/group-buying-recommendation}.}

In order to adapt the dataset to collaborative filtering methods and social recommendation methods,
\rev{group-buying behavioral records should be converted into pure user-item interactions.}
\rev{There are two conversion methods: (1) only retain the initiator-item interactions or (2) treat both initiator-item and participant-item interactions as pure user-item interactions.}
\rev{Following experiments will show which method is better.}
\rev{For group recommendation methods, we consider each user and those who do group buying with him/her as a group; then each successful behavior is regarded as one activity of its initiator's group.}
However, we keep the parts for testing consistent in the original dataset and its variants above to make performance comparison equitable.
\rev{The exception is that we replace each user with the group corresponding to the user as the input of group recommendation methods when testing.}

\begin{table}[t]
	\centering
	\caption{\rev{Statistics of the dataset.}}
	\label{tab::stat}
	\begin{tabular}{c|c|c|c}
		\hline
		\multicolumn{2}{c|}{\#Users} & \multicolumn{2}{c}{190,080} \\ \hline
		\multicolumn{2}{c|}{\#Items} & \multicolumn{2}{c}{30,782} \\ \hline
		\multicolumn{2}{c|}{\#Social Interactions} & \multicolumn{2}{c}{748,233} \\ \hline
		\multirow{2}{*}{\#Group-buying Behaviors} & \multirow{2}{*}{932,896} & \#Successful & 721,605 \\ \cline{3-4}
		~& ~& \#Failed & 211,291 \\ \hline
	\end{tabular}
\vspace{-0.4cm}
\end{table}

\subsubsection{Evaluation Protocols}
Following existing works~\cite{ncf,he2019joint,chen2020sequence,gao2019neural}, the widely used \textit{leave-one-out} evaluation protocol is applied during the evaluation phase and the dataset is divided into the training set and the testing set.
From the training set, we also randomly select one record for each user as the validation set to tune hyper-parameters. For the testing and validation set, we randomly sample 999 items which have not been interacted by the user given a user as initiator, and then all methods rank one test item and these sampled items.
To report the ranking performance of all methods, we adopt two ranking metrics, \textit{Recall} and \textit{NDCG}, widely used in recommendation~\cite{zheng2020price,madisetty2019event,khawar2019modeling}:
\begin{itemize}[leftmargin=*]
	\item \textbf{Recall@K}: Recall measures whether the test item is present in the top-K item ranking list (1 if the answer is yes else 0).
	\item \textbf{NDCG@K}: \textit{Normalized Discounted Cumulative Gain} (NDCG) complements Recall by evaluating the location where the test item appears rather than only considering whether it appears.
\end{itemize}

It should be noted that the reported performance is the averaged value for all users.

\begin{table*}[t]
	\centering
	\caption{Overall performance comparison on our collected Beibei Dataset}
	\label{tab::perf}
	\begin{tabular}{c|c|c|c|c|c|c|c|c|c}
		\hline
		\textbf{Category} & \textbf{Method} & \textbf{Recall@3} & \textbf{Recall@5}  & \textbf{Recall@10}  & \textbf{Recall@20} & \textbf{NDCG@3}  & \textbf{NDCG@5} & \textbf{NDCG@10}  & \textbf{NDCG@20} \\ \hline
		\multirow{4}{*}{CF} & \rev{\textbf{MF}(oi)}      & \rev{0.0762}& 	\rev{0.1055} &	\rev{0.1567} &	\rev{0.2219} &	\rev{0.0590} &	\rev{0.0710} &	\rev{0.0875} &	\rev{0.1039}    \\ \cline{2-10}
		~& \textbf{MF}      & 0.1086 & 0.1456 & 0.2106 & 0.2886 & 0.0847 & 0.0999 & 0.1208 & 0.1405  \\ \cline{2-10}
		~& \textbf{NCF}  &0.1231 & 0.1640 & 0.2327 & 0.3142 & 0.0961 & 0.1129 & 0.1351 & 0.1556 \\ \cline{2-10}
		~& \textbf{NGCF}         &0.1171 & 0.1556 & 0.2190 & 0.2958 & 0.0922 & 0.1080 & 0.1284 & 0.1478 \\ \hline
		\multirow{2}{*}{S}& \textbf{SocialMF}       &0.1135 & 0.1532 & 0.2202 & 0.3013 & 0.0889 & 0.1051 & 0.1268 & 0.1472\\ \cline{2-10}
		~& \textbf{DiffNet}         & 0.1249 & 0.1664 & 0.2332 & 0.3153 & 0.0981 & 0.1151 & 0.1366 & 0.1573\\ \hline
		\multirow{2}{*}{G}& \textbf{AGREE}  & 0.1036 & 0.1441 & 0.2097 & 0.2806 & 0.0798 & 0.0964 & 0.1175 & 0.1355  \\ \cline{2-10}
		~& \textbf{SIGR}  & 0.1038 & 0.1405 & 0.2034 & 0.2809 & 0.0806 & 0.0956 & 0.1159 & 0.1354 \\ \hline
		\multirow{2}{*}{GB}& \rev{\textbf{GBMF}} & \rev{0.1262} &	\rev{0.1678} &	\rev{0.2350} &	\rev{0.3141} &	\rev{0.0991} &	\rev{0.1162} &	\rev{0.1379} &	\rev{0.1578} \\ \cline{2-10}
		~ & \textbf{GBGCN}      & \bf 0.1341 & \bf 0.1756 & \bf 0.2444 & \bf 0.3237 & \bf 0.1064 & \bf 0.1234 & \bf 0.1456 & \bf 0.1656  \\ \hline
		& \rev{\textbf{Improvement}} &\rev{6.29\%} & \rev{4.67\%}&	\rev{4.04\%}&	\rev{2.69\%}&	\rev{7.36\%}&	\rev{6.26\%}&	\rev{5.64\%}&	\rev{4.97\%} \\ \hline
	\end{tabular}
\end{table*}

\subsection{Overall Performance (RQ1)}\label{sec::rq1}
\subsubsection{Baselines}
We compared our GBGCN with four categories of baseline methods, including collaborative filtering methods, social recommendation methods, group recommendation methods \rev{and one group-buying recommendation method}.

Collaborative filtering methods refer to those models that can only utilize user-item interaction data.
The compared collaborative filtering methods are introduced as follows:
\begin{itemize}[leftmargin=*]
	\item \textbf{MF}~\cite{MF}.
	\textit{Matrix Factorization} (MF) is a widely used collaborative filtering model which exploits user implicit feedback to extract user preference and item feature.
	It adopts BPR as the loss function and samples a number of negative items with a certain ratio.
	Both positive and negative items are fed into the model for training in a certain proportion.
	We tune the learning rate and regularizer and then report the performance of the best model on the validation set.
	\item \textbf{NCF}~\cite{ncf}.
	\textit{Neural network-based Collaborative Filtering} (NCF) is the state-of-the-art neural solution for recommendation tasks with implicit feedback.
	It ensembles Generalized Matrix Factorization (GMF) and Multi-Layer Perceptron (MLP) to model non-linearities in user-item interactions.
	We tune its hyper-parameters according to the related settings in the paper.
	\item \textbf{NGCF}~\cite{NGCF}.
	\textit{Neural Graph Collaborative Filtering} (NGCF) is a recent recommendation framework based on graph neural networks.
	It uses embedding propagation on the user-item bipartite graph to model high-order connectivity and extract collaborative signals.
	We optimize this model and tune it following the paper.
\end{itemize}
The compared social recommendation methods are as follows:
\begin{itemize}[leftmargin=*]
	\item \textbf{SocialMF}~\cite{jamali2010matrix}.
	SocialMF is an MF-based model for recommendation with social networks, which force the preference of a user to be closer to the average preference of the user's social relations.
	BPR is also adopted as the loss function.
	We tune its learning rate and regularizer similarly with MF.
	\item \textbf{DiffNet}~\cite{DiffNet}.
	Diffnet is a state-of-the-art social recommendation model based on GCN.
	It leverages social relations and stimulates the recursive social influence propagation process on social networks for better embedding modeling.
	We tune the diffusion depth, learning rate, and coefficient of regularization term before reporting the best performance.
\end{itemize}
The compared group recommendation methods are as follows:
\begin{itemize}[leftmargin=*]
	\item \textbf{AGREE}~\cite{cao2018attentive}.
	\textit{Attentive Group Recommendation} (AGREE) is based on NCF and the attention mechanism.
	It adopts an attention mechanism to obtain group representation and learns the interaction function between group and item by neural networks.
	We follow the pre-training pipeline in the paper to optimize the model on the variant of Beibei dataset for group recommendation.
	\item \textbf{SIGR}~\cite{yin2019social}.
	\textit{Social Influence-based Group Recommender} (SIGR) is composed of a bipartite graph embedding model that leverages user-item graph and group-item graph and an attention mechanism used for learning user social influence in different groups.
	We follow the paper and try our best to tune it.
\end{itemize}
\rev{The compared group-buying recommendation method is as follows}:
\begin{itemize}[leftmargin=*]
	\item \rev{\textbf{GBMF}.}
	\rev{\textit{Group-Buying Matrix Factorization} (GBMF) is a reasonable but intuitive solution for our problem compared with GBGCN.}
	\rev{The baseline uses dot-based similarity function to calculate user's interest for the given item, computes the average score of the potential initiator's friends and then weighted sum is applied to estimate the final score and balance initiator and participants interest.}
	\rev{BPR is adopted as the loss function.}
	\rev{We tune the baseline like GBGCN.}
\end{itemize}

We implement the baseline methods and our GBGCN model in PyTorch\footnote{https://pytorch.org} with Deep Graph Library (DGL)\footnote{https://www.dgl.ai}.
We set the embedding size of all compared methods to be 32 to balance memory efficiency and model capacity, which is a common setting in existing works on graph representation learning based recommendation~\cite{wang2019neural,wang2019kgat}.
Our main experiments also demonstrate that such embedding size has enough representation ability.

\subsubsection{Parameter Settings}
We tune all the hyper-parameters of our method on the validation set.
We initialize parameters with the widely used Xavier initialization method~\cite{pmlr-v9-glorot10a}.
When training, we set the negative sampling ratio as 1:1 following~\cite{Zheng2020PriceawareRW} and construct mini-batches with a size of 4096 to balance training efficiency and memory usage.
To optimize the GBGCN model better, we follow the strategy described in~\cite{cao2018attentive,ncf}, employ Adam in the pre-training stage, and use the vanilla SGD in fine-tuning stage to avoid the problem of loss of momentum information.
Besides, we search the Adam's learning rate within \{1e-2, 1e-3, 1e-4, 1e-5\} and search the SGD's in \{10, 3, 1, 0.3\}.
In addition, we search the coefficients of regularization term in \{1e-1, 3e-2, 1e-2, 3e-3, 1e-3, 3e-4, 1e-4, 3e-5, 1e-5\}.
$\alpha$, the coefficient to balance the importance of initiator and participants, is searched at intervals of 0.1 within the range of 0.1 to 0.9.
We also tune the loss coefficient $\beta$ in \{0.01, 0.02, 0.05, 0.1, 0.2,0.5\}. The number of layers $L$ is set to 2.
For each model, we train it for 500 epochs, and we save the model that has the best performance on the validation set for evaluation.

Firstly, we compare the overall performance of GBGCN and baseline methods on the top-K group-buying recommendation task.
In order to fully demonstrate the ability of the methods in different segments, K is set to \{3, 5, 10, 20\}.
It is another reason for relatively small K that the length of ranking list is 1000 according to the evaluation protocols described above.
We tune all the hyper-parameters for each method carefully and report their best preformance on the real-world dataset in Table~\ref{tab::perf}. 
In the category column in Table~\ref{tab::perf}, CF, S, G, GB are the abbreviations of collaborative filtering, social recommendation, group recommendation and group-buying recommendation, respectively.
\rev{Because two conversion methods are proposed above, we use \textit{oi} to mark the model in Table~\ref{tab::perf} that adopts the first conversion method, which retains initiator-item interactions only and do not mark models using the second one to keep the table concise.}
From these results, we obtain several observations as follows:

\begin{itemize}[leftmargin=*]
	\item \textbf{Our GBGCN method outperforms all the baseline methods on the real-world dataset significantly.}
	Benefiting from the directed heterogeneous graphs and multi-view information propagation, GBGCN is capable of modeling the multi-role feature and complex interactions in group buying.
	The fine-grained loss function helps GBGCN fully mine information from the dataset.
	Compared to all baseline methods, GBGCN obtains the best performance in terms of Recall@K and NDCG@K and outperforms the best baseline by 2.69\%-\rev{6.29\%} in Recall and \rev{4.97\%-7.36\%} in NDCG. The p-value of significance tests is less than 0.05, which demonstrates the improvement is stable and significant.
	It is noted that GBGCN performs better with smaller K, which is more meaningful for application scenarios.
	Such improvements on various metrics demonstrate GBGCN's effectiveness. 
	\item \rev{\textbf{The performance of MF with only initiator-item interactions is much lower than that with both initiator-item and participant-item interactions.}}
	\rev{As we can observe, MF with extra data achieves better prediction performance.}
	\rev{This phenomenon indicates that collaborative filtering and social recommendation models should be fed as much data as possible to model user preference and item feature precisely.}
	\rev{The same phenomenon also appears in experiments of other models so that we do not show their performance.}
	\item \textbf{As collaborative filtering methods, NCF generally achieves better performance than NGCF.}
	According to ~\cite{NGCF}, NGCF outperforms NCF on most item recommendation tasks.
	However, it is multiplex for the relations between user and item in our problem.
	NGCF is designed for exploring collaborative filtering signals in the user-item bipartite graph and might fails to fully model the multiplex relations while NCF is capable of describing them by MLP. 
	\item \textbf{As group recommendation methods, the performance of AGREE and SIGR are both poor, even worse than MF.}
	In group recommendation problem, the recommender is required to recommend items for a fixed group, and users have no role difference.
	To measure user's influence in a group, the two methods both introduce attention mechanism to aggregate user embeddings as group embeddings while attention mechanisms do not work due to the data sparsity problem of the real-world dataset.
	More importantly, users have dynamic roles in group buying, which is hard for group recommendation methods to handle.
	The selection of loss function may be another reason.
	AGREE adopts a regression-based pairwise loss, which forces the margin of the prediction of observed interaction and unobserved one to be 1.
	SIGR adopts \textit{logloss} to classify positive and negative items.
	Maybe it is relatively weaker for the two loss funcions to obtain information from negative sampling than BPR.
	\item \rev{\textbf{The performance of GBMF is the best among all the baseline methods in terms of most metrics.}}
	\rev{As an intuitive but strictly suitable baseline for our problem, the performance of GBMF indicates that targeted design must be made for new problems and it is unsuitable to adapt methods which are built for different problems.}
	\rev{Meanwhile, the performance comparison between GBGCN and GBMF indicates the benefits of the directed heterogeneous graphs, multi-view information propagation and the fine-grained loss function proposed in GBGCN.}
\end{itemize}

To summarize, the extensive comparisons on the real-world dataset verify that our GBGCN method is effective and suitable for our problem.

\begin{table}[t]
	\centering
	\caption{\rev{Time efficiency comparison.}}
	\label{tab::time}
	\begin{tabular}{c|c|c}
		\hline
		\textbf{Methods} & \textbf{Training Time} (sec/epoch) & \textbf{Testing Time} (sec/epoch) \\ \hline
		\textbf{MF(oi)}& 2.99& 4.74 \\ \hline
		\bf MF& 3.65& 4.75 \\ \hline
		\bf NCF& 3.83& 4.47 \\ \hline
		\bf NGCF& 5.68& 4.87 \\ \hline
		\bf SocialMF& 5.27& 4.83 \\ \hline
		\bf DiffNet& 4.77& 4.55 \\ \hline
		\bf AGREE& 17.25& 15.25 \\ \hline
		\bf SIGR& 58.29& 8.56 \\ \hline
		\bf GBMF& 31.68& 54.34 \\ \hline
		\bf GBGCN& 56.28& 88.36 \\ \hline
	\end{tabular}
	\vspace{-0.3cm}
\end{table}

\subsection{\rev{Time Efficiency Comparison}}
\rev{In this section, we conduct further experiments to study how time efficient our GBGCN and baselines are.}
\rev{All these experiments are run on the same single Linux server with one NVIDIA TITAN Xp and DGL is adopted to speed up all GCN-based models.}
\rev{Table~\ref{tab::time} reports the time efficiency results (measured in wall-clock time).}

\rev{From the results, we can observe that collaborative filtering and social recommendation baselines all achieve good training and testing time efficiency due to relatively simple input data format.}
\rev{However, group recommendation and group-buying recommendation models spend more time on training and testing.}
\rev{The reason is that these models all consider user's friends or group members in forwarding phase but the number of each user's friends or each group's members is different.}
\rev{The inconsistent number of elements greatly impairs the parallelism of these models, resulting in a slower running speed.}

\subsection{Impact of Multi-view Design (RQ3)}\label{sec::rq3}
In our GBGCN, we generate two sets of embeddings for initiator view and participant view separately by in-view propagation and cross-view propagation to model user's and item's different characteristics in different views and overcome the first major challenge mentioned in the introduction. There are two intuitive questions about whether the multi-view design helps in GBGCN, and how does it affect the performance?

To answer them, we conduct ablation experiments on three degenerative models of GBGCN.
Without reducing the capacity of the model, we simply replace the two sets of embeddings at the output of each propagation layer with their average pooling.
To be specific, if we remove the user roles in a 2-layer GBGCN, $u_{m,i}^{(l)}$ and $u_{m,p}^{(l)}$ will be replaced by $\frac{u_{m,i}^{(l)}+u_{m,p}^{(l)}}{2}$ for $l=1,2$ after each propagation, and the treatment of $u_{m,i}^{\{1\}}$ and $u_{m,p}^{\{1\}}$ is also consistent.
The ablation manners for removing only item roles and removing both user and item roles are similar.
We adopt the same evaluation methods with the above experiments and report the performance comparison in Table~\ref{tab::role}.
We can observe that the recommendation performance becomes poor after removing either item roles or user roles.
Meanwhile, it is intuitive that removing both user and item roles makes the performance worse than removing only one part.
To conclude, the experimental results of the ablation study demonstrate the necessity of the multi-view design.

\begin{table*}[t]
	\centering
	\caption{Impact of Multi-view Design}
	\label{tab::role}
	\begin{tabular}{c|c|c|c|c|c|c|c|c}
		\hline
		\textbf{Methods} & \textbf{Recall@10} & \textbf{Improve.}&\textbf{Recall@20} &\textbf{Improve.} &\textbf{NDCG@10} & \textbf{Improve.}& \textbf{NDCG@20}&\textbf{Improve.} \\ \hline
		\textbf{GBGCN} & 0.2444 &-& 0.3237 &-& 0.1456 &-& 0.1656&- \\ \hline
		\textbf{Without Item Roles} & 0.2422 &-0.93\%& 0.3226 &-0.36\% & 0.1439 &-1.17\% & 0.1642 &-0.87\% \\ \hline
		\textbf{Without User Roles} & 0.2430 &-0.59\% & 0.3218 &-0.59\% & 0.1447 &-0.66\% & 0.1646 &-0.65\% \\ \hline
		\textbf{Without Item and User Roles} & 0.2408 & -1.49\% & 0.3189 &-1.49\%& 0.1439 &-1.18\% & 0.1636 &-1.21\% \\ \hline
	\end{tabular}
\end{table*}

\begin{figure*}[t]
	\begin{center}
		\mbox{
			\subfloat{\includegraphics[width=0.222\linewidth]{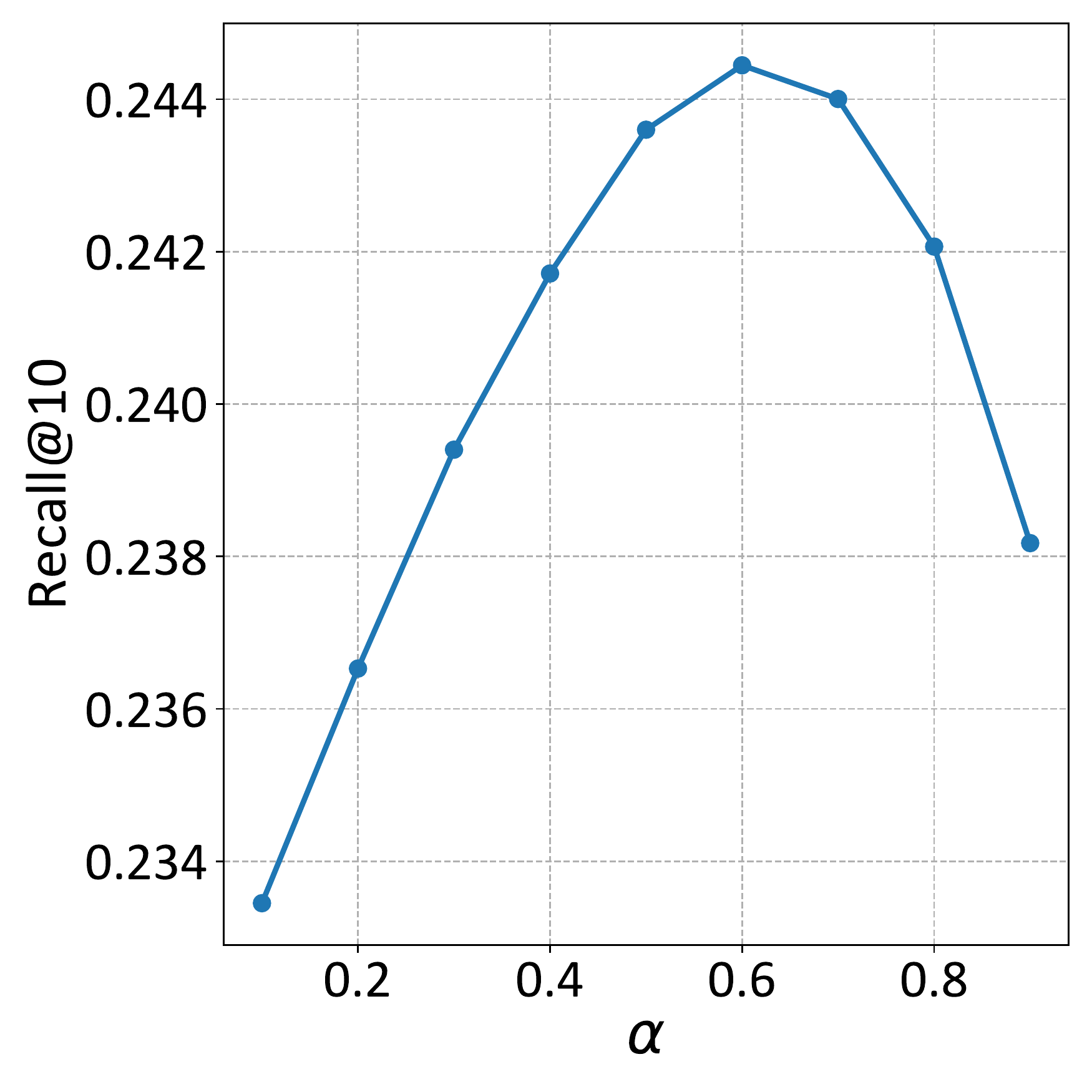}}
			\subfloat{\includegraphics[width=0.222\linewidth]{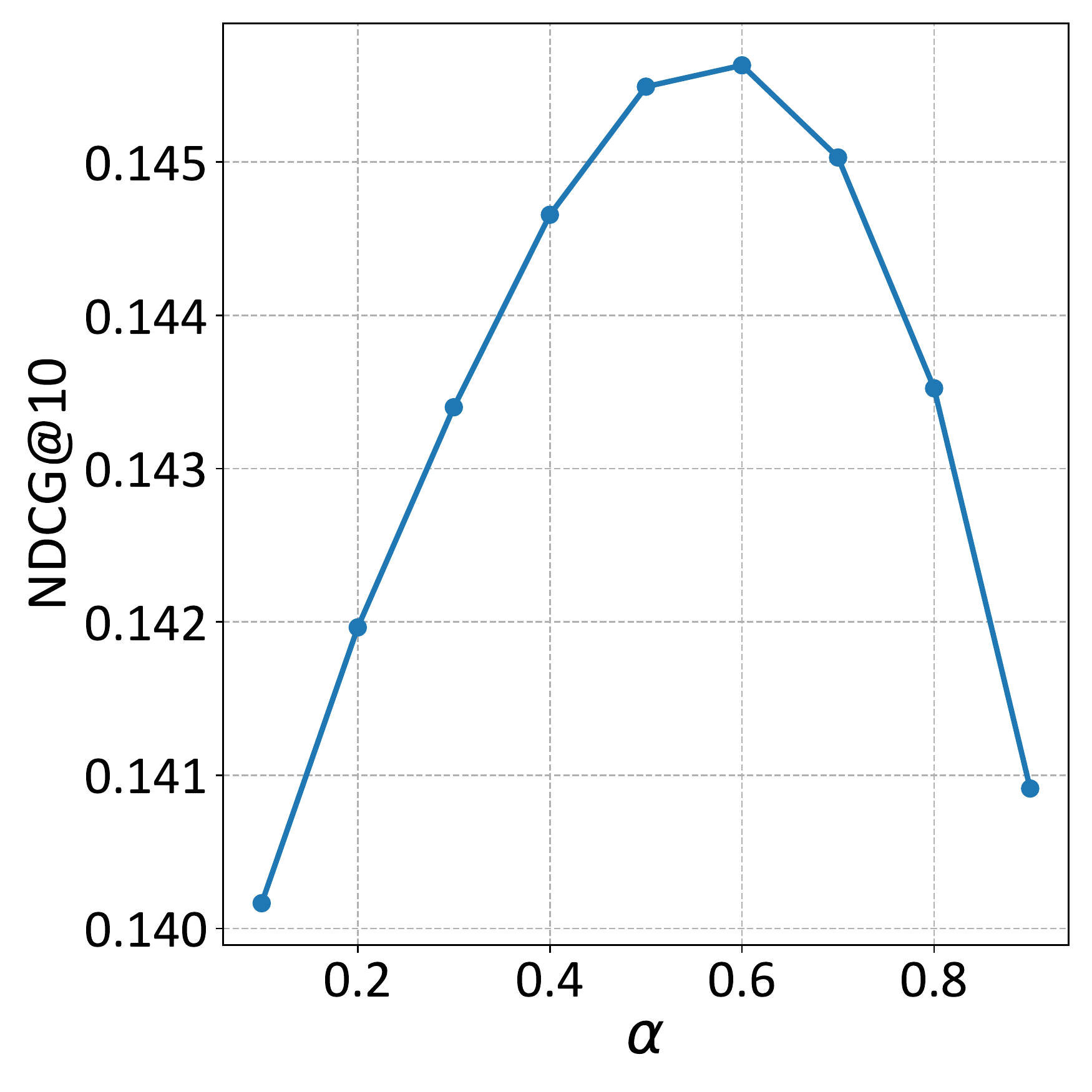}}
			\subfloat{\includegraphics[width=0.222\linewidth]{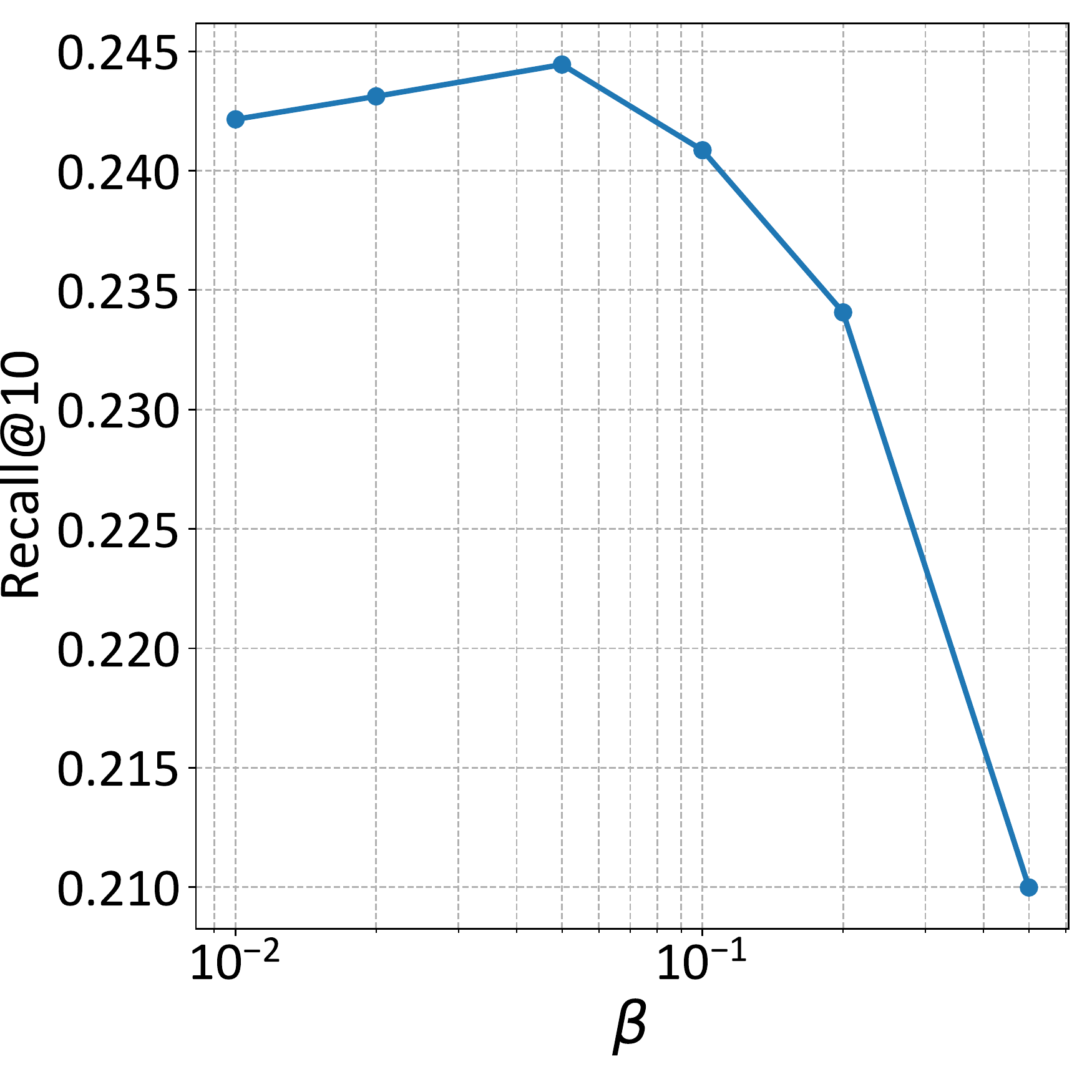}}
			\subfloat{\includegraphics[width=0.222\linewidth]{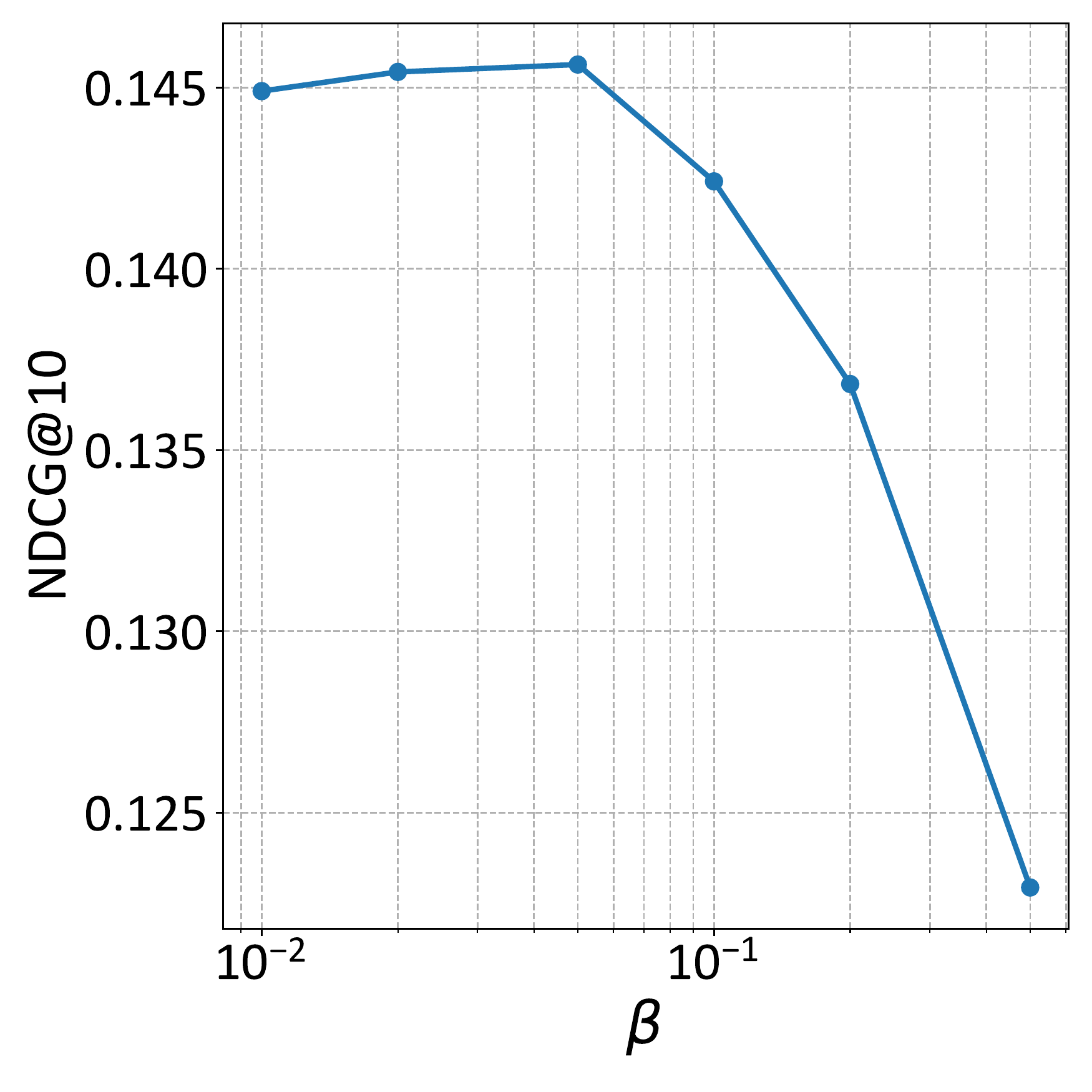}}
		}
	\end{center}
	\vspace{-0.3cm}
	\caption{Testing performance with different role coefficient $\alpha$ and different loss coefficient $\beta$.\label{fig::rq4}}
		\vspace{-0.3cm}
\end{figure*}

\begin{figure*}[t]
	\centering
	\subfloat[]{\includegraphics[width=0.222\linewidth]{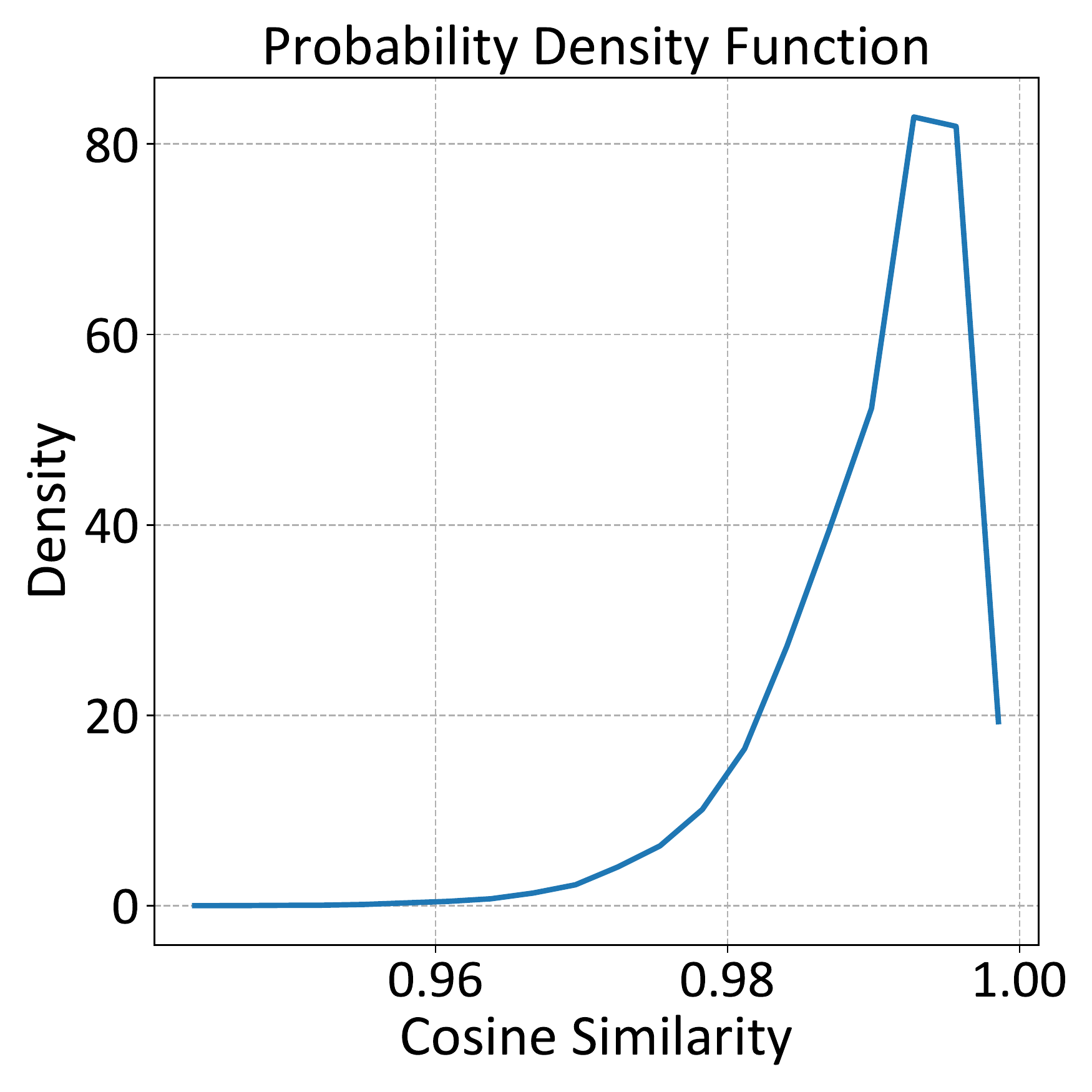}}
	\subfloat[]{\includegraphics[width=0.222\linewidth]{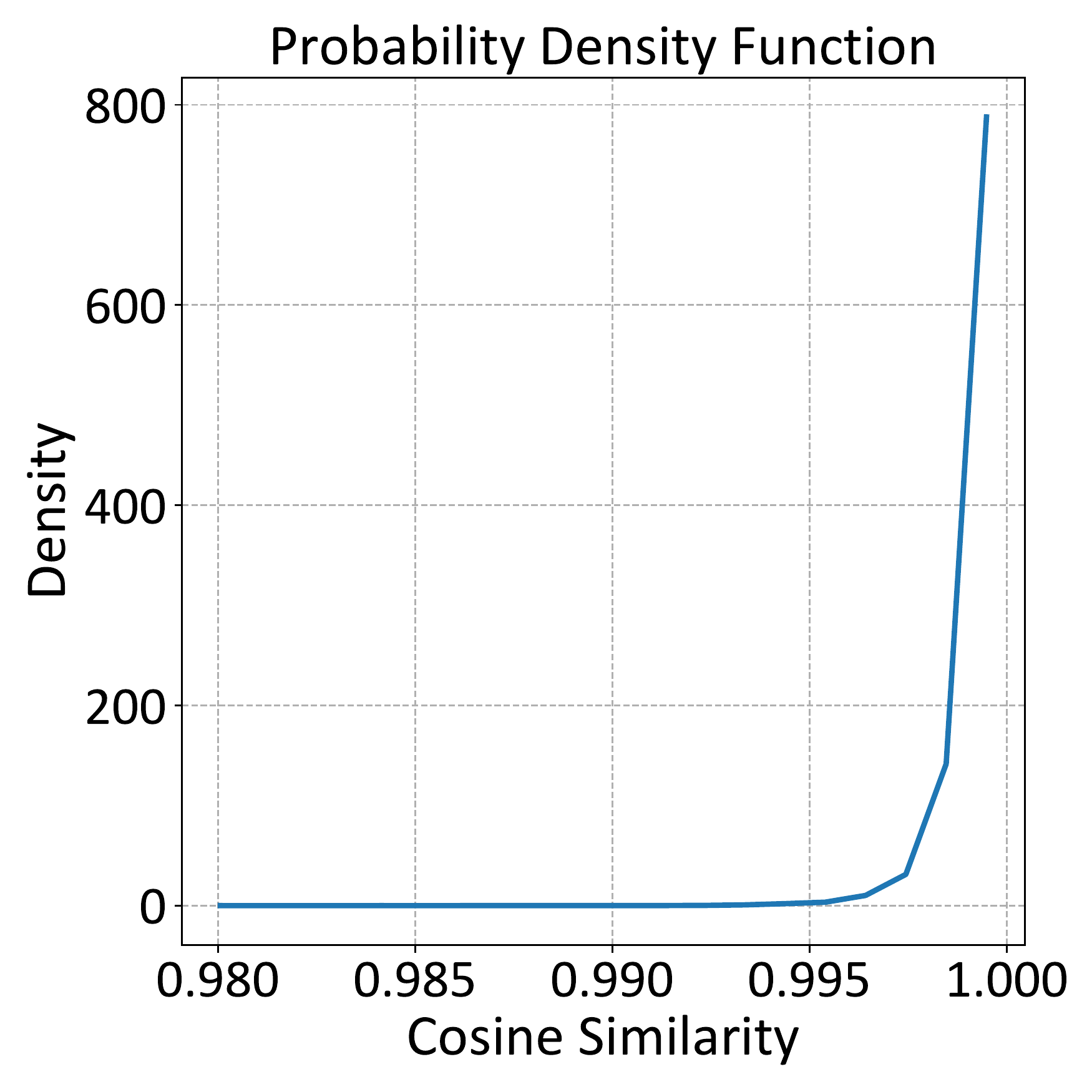}}
	\subfloat[]{\includegraphics[width=0.222\linewidth]{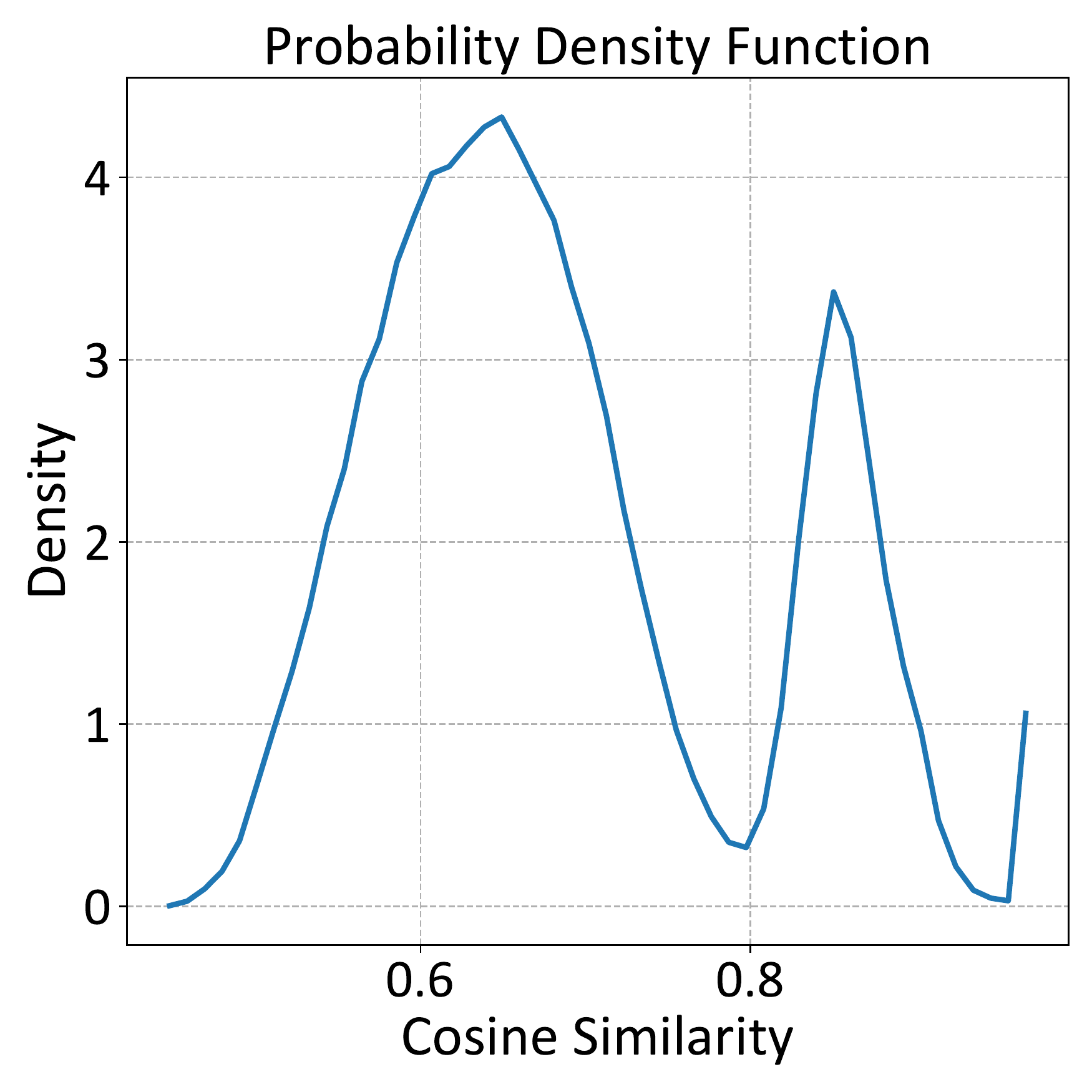}}
	\subfloat[]{\includegraphics[width=0.222\linewidth]{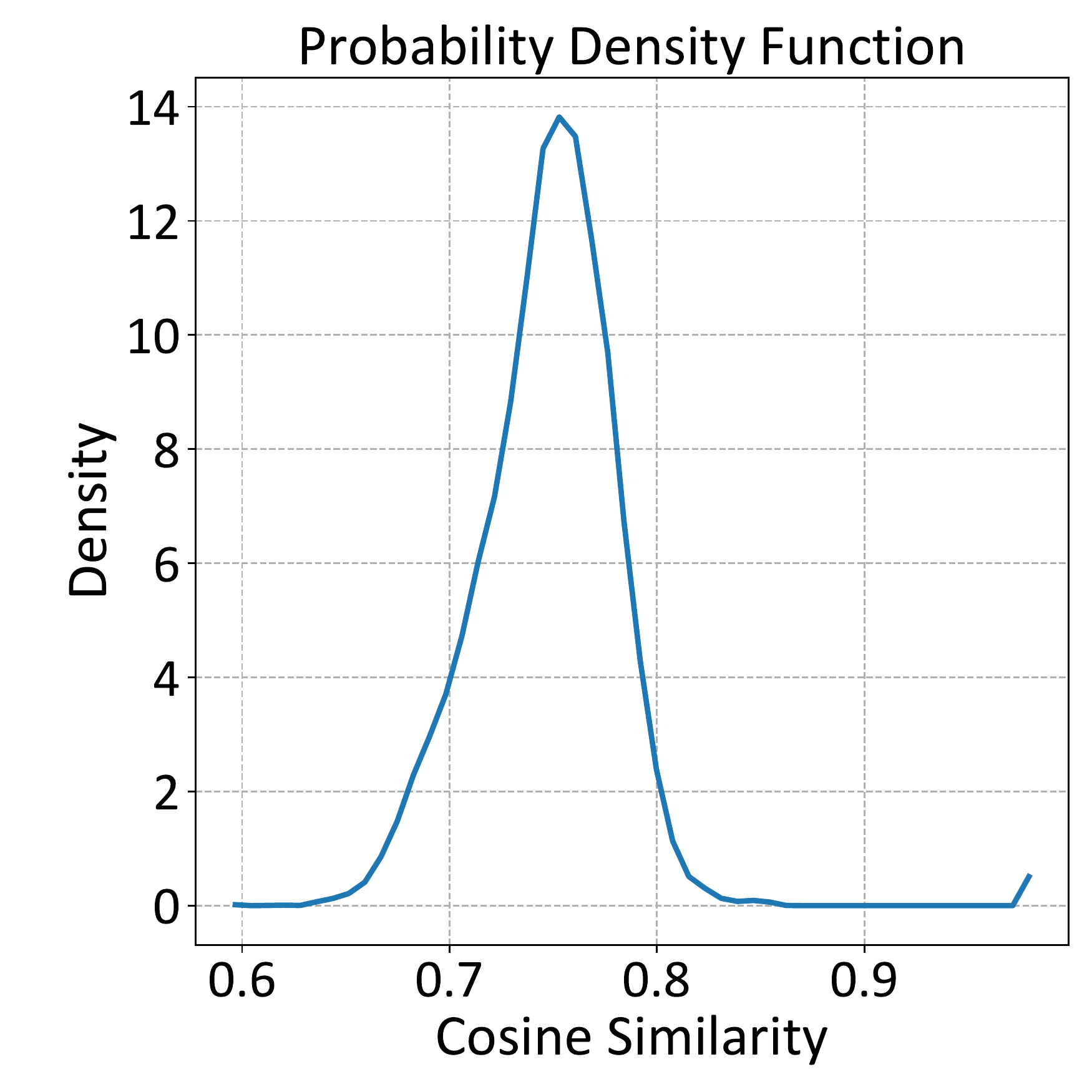}}
	\vspace{-0.3cm}
	\caption{Probability density function (PDF) of cosine similarity between embeddings  in initiator view and participant view. (a) shows PDF of cosine similarity of the two sets of user's in-view propagation generated embeddings; (b) shows PDF of cosine similarity of the two sets of item's in-view propagation generated embeddings; (c) shows PDF of cosine similarity of the two sets of user's cross-view propagation generated embeddings; (d) shows PDF of cosine similarity of the two sets of item's cross-view propagation generated embeddings.\label{fig::rq5-stat}}
\end{figure*}

\begin{figure}[thb]
	\begin{center}
		\mbox{
			\subfloat{\includegraphics[width=0.80\linewidth]{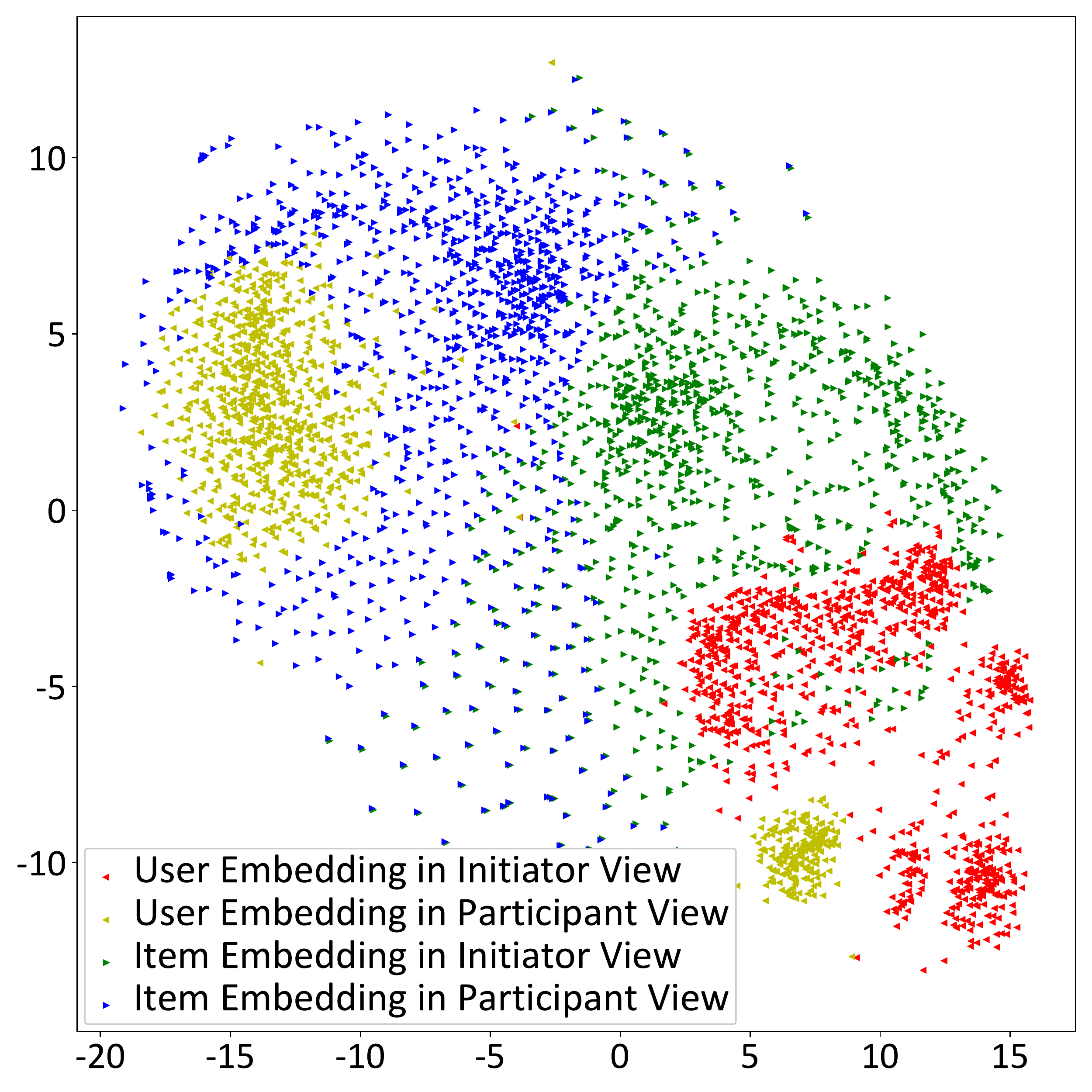}}
		}
	\end{center}
	\caption{The visualization of user and item embedding in initiator and participant view by t-SNE.  (Best view in color)}
	\label{fig::rq5-tsne}
\end{figure}

\subsection{Hyper-parameter Study (RQ4)}\label{sec::rq4}

For our GBGCN method, there are two significant hyper-parameters that are closely related to performance.
In this section, we study how much these hyper-parameters affect prediction performance and find out the best hyper-parameters.
We investigate the role coefficient $\alpha$ mentioned in Equation~\ref{eqn::pred}, followed by the loss coefficient $\beta$ in Equation~\ref{eqn::loss-failed}.

\subsubsection{Role Coefficient}
The role coefficient $\alpha$, which is used in the prediction function, balances the influence of initiator and participant candidates for a successful group buying.
A larger $\alpha$ implies that users on e-commerce websites tend to consider their friends' interest when launching deals.
On the contrary, a smaller $\alpha$ means that users prioritize their own needs, and it is relatively easy to find enough participants to some extent.

To study how the coefficient affects the prediction performance, we present the Recall@10 and NDCG@10 performance in Figure~\ref{fig::rq4}.
We can observe that a biased value of $\alpha$ worsen the prediction performance, and simply averaging is also not the best solution.
Our method GBGCN achieves its best performance on the real-world dataset with $\alpha=0.6$.

\subsubsection{Loss Coefficient}

We make use of the loss coefficient $\beta$ to adjust the strength of belief for strong-negative feedback, failed group-buying behaviors.
Actually, the reasons why a group failed are complicated. 
In addition to the weak social influence and friends' interest, there are also possible reasons such as the user's wrong operations, not sharing after purchase, etc.
Therefore, the strong-negative feedback is not completely reliable.

To study what value of $\beta$ is most beneficial to GBGCN, we search $\beta$  in \{0, 0.01, 0.02, 0.05, 0.1, 0.2,0.5\} and present the Recall@10 and NDCG@10 performance in Figure~\ref{fig::rq4} except $\beta=0$.
We can observe that appropriate value of $\beta$ really helps GBGCN on the dataset. The best $\beta$ value is 0.05.
\rev{When $\beta=0$, the double-pairwise loss degenerates to the standard pairwise loss and Recall@10 is 0.2401 and NDCG@10 is 0.1421, which is worse than our method. This result indicates that our double-pairwise loss is better than the standard pairwise loss.}

\subsection{Embedding Analysis (RQ5)}\label{sec::rq5}

In GBGCN, we introduce the multi-view design to model multiple roles of users in group buying.
We conduct ablation studies about this design and then verify its necessity in Section~\ref{sec::rq3}.
We argue that the two sets of embeddings generated by multi-graph convolution networks contain specific information on the corresponding view.
To verify it and obtain some insight from the embeddings, we
(1) compute the probability density function (PDF) for cosine similarity of in-view propagation generated embeddings and cross-view propagation generated embeddings separately between initiator view and participant view for each entry and
(2) visualize the final embeddings of randomly sampled 1000 items and 1000 users by a widely used method t-SNE~\cite{maaten2008visualizing}.

The PDF curves are shown in Figure~\ref{fig::rq5-stat}.
We can observe that for in-view propagation generated embeddings, the embeddings of items are basically consistent in different views, and the embeddings of users show a certain difference.
This phenomenon indicates that item characteristics reflected in initiator-item interactions and participant-item interactions are close, while users' interests start to diverge.
For cross-view propagation generated embeddings,  embeddings in the two views diverge obviously, which means that cross-view propagation and FC layers succeed in capture view-specific information.

The visualization is shown in Figure~\ref{fig::rq5-tsne} also supports the insight above.
Ignoring the clusters in the lower right corner, which might be users with few interactions in the corresponding view, the nodes in the figure can be seen as two parts, embeddings in initiator view on the right side and embeddings in participant view on the left side. The visualization demonstrates the ability of GBGCN to model the difference between the two views at the embedding level.

In conclusion, extensive experiments on the real-world dataset verify the efficacy of our proposed GBGCN. Further studies demonstrate that GBGCN succeeds in generating proper representation for both initiator view and participant view by the multi-view design.

%% file: 5.related.tex
\section{Related Work}\label{sec::related}
\vspace{-0.4cm}
\rev{
\subsection{Group Buying and Group Recommendation}
Group buying~\cite{anand2003group} is generally defined as that a group of people purchase the same product together. Actually, there are two kinds of forms of group buying in today’s online e-commerce services. At the earlier group-buying~\cite{bai2019hybrid,hu2014multicriteria,chen2015we}, platforms help to organize a team consisting hundreds or even thousands users—who do not know each other in most times—to purchase items together from sellers. Recently, a new kind of group-buying has become very popular, such as Pinduoduo in China. Specifically, users can serve as the group founder, and invite their friends to join the group. As a result, the users in group of the new kind of group buying are always connected in social networks, which make it completely different from the previous one.
}

Another close-related topic is group recommendation~\cite{amer2009group} is to recommend one item to a group of users. \rev{All users in the group interact with \textit{one} item together}, and groups are always pre-defined and constant. Typical applications of group recommendation consist of travel recommendation, restaurant recommendation, etc. 
Early approaches~\cite{baltrunas2010group,liu2012exploring,quintarelli2016recommending,yuan2014generative,christensen2016social,hu2014deep}  model the interest of each user belonging to one group and then design some strategies, such as average,
least misery, maximum satisfaction, etc., to fuse the interests together. 
Cao~\textit{et al.}~\cite{cao2018attentive} propose to utilize neural collaborative filtering with attention mechanisms to model the user interest and group interest at the same time, which is further extended by the same authors with combining social network~\cite{cao2019social}.
Yin~\textit{et al.}~\cite{yin2019social} consider the social relation in pre-defined groups, and introduce social-aware constraint into the preference learning in groups. 
In our problem of group-buying recommendation for social e-commerce, the group is formed dynamically from the social network, and there is no pre-defined group at all, which is very different compared with group recommendation tasks.

\rev{
\subsection{Social Influence and Social Recommendation}
Social influence~\cite{cialdini2004social,jackson2010social} studies how friends affect users' behaviors or decisions and has wide applications in various areas~\cite{yin2019social,feng2018inf2vec,wang2016distance,kim2013scalable}.
For recommender systems, social influence is introduced by the \textit{social-trust} assumption that users trusts their friends' decisions and will have similar preferences with friends. The recommendation models that leverage social influence with social-network data is defined as social recommendation~\cite{socialrecommendationreview}.
} 
Based on the social-trust assumption, early works used regularization techniques to limit the distance of embedding vectors in latent space between a user and her friends~\cite{jamali2010matrix,guo2015social,soreg,soregbpr,wang2017item}.
Some works assume a user's decision should be a combination of her own characteristics and her trusted friends' recommendations~\cite{ste,mtrust,icde-ensemble}.
Some other works~\cite{sorec,localbal,lin2019cross} consider the recommendation and social relation prediction as two tasks and approach the problem with multi-task learning.
Yu~\textit{et al.}~\cite{yu2019generating} proposed to utilize generative adversarial networks to identifying reliable friends from observed and unobserved social networks.
Wu~\textit{et al.}~\cite{DiffNet} proposed to use embedding propagation on the social network to capture the social-trust effect. Fan~\textit{et al.}~\cite{fan2019graph} further combine embedding propagation with attention networks to model friends' various influence.
Xu~\textit{et al.}~\cite{xu2019relation} considered the recommendation scenario in social e-commerce when users serve as selling agents and leverage multi-hop graph convolutional networks to learn the social influence.

Different from social recommendation, the aim of our problem is a group of users, rather than a single user. The only same point is that both two takes take use of social networks to help learn user preferences and users' social influence.

\subsection{Graph Convolutional Networks for Recommendation}
Graph convolutional network~\cite{kipf2016semi} has become a new state-of-the-art approach of graph representation learning.
The core idea of GCN is to use embedding propagation to extract high-order structural knowledge and node attributes at the same time. 
~\cite{GCMC} first applied GCN to collaborative filtering recommendation, which can be considered as extension to SVD++~\cite{MF}. 
Ying~\textit{et al.}~\cite{ying2018graph} further extended GCMC to implicit recommender systems with a neighbor-sampling technique to reducing computation cost. Wang~\textit{et al.}~\cite{wang2019neural} further adjusted the design of the propagation and prediction layer, which is a state-of-the-art GCN based collaborative filtering method.
GCN also achieved big success on content-based recommendation, such as session-based recommendation~\cite{wu2019session}, knowledge-based recommendation~\cite{wang2019kgat}, price-aware recommendation~\cite{zheng2020price}, multi-behavior recommendation~\cite{jin2020multi}, bundle recommendation~\cite{chang2020bundle}, etc.
However, these existing GCN-based recommendation models are designed for collaborative filtering tasks or specific tasks with unique forms of data. Thus, they cannot be applied to our problem of group-buying recommendation.

%% file: 6.conclusion.tex
\section{Conclusion and Future Work}\label{sec:conclusion}

In this work, we take the pioneer step to approach the problem of group-buying recommendation for social e-commerce. We propose to construct directed hetegenous graphs to represent group-buying behaviors, including launching and joining groups, and social network data. We develop a graph convolutional network method named GBGCN, which combines inner-view and cross-view embedding propagation with a double-pairwise loss function. We collect a real-world group-buying dataset and conduct extensive experiments. The results demonstrate that our GBGCN can achieve the best recommendation performance compared with baselines with significant improvements.
As for future work, we plan to study the data sparsity issue and conduct A/B test to further evaluate our GBGCN model.